\documentclass[aps,prd,twocolumn,preprintnumbers,superscriptaddress,floatfix]{revtex4}

\usepackage{multirow, graphicx,amssymb,url,mathrsfs,amsmath}
\usepackage{eucal,wrapfig,boxedminipage,setspace,subfigure}
\usepackage{amsxtra,amstext,latexsym,dsfont}
\usepackage[colorlinks]{hyperref}

\def\IR{{\hbox{{\rm I}\kern-.2em\hbox{\rm R}}}}
\def\IB{{\hbox{{\rm I}\kern-.2em\hbox{\rm B}}}}
\def\IN{{\hbox{{\rm I}\kern-.2em\hbox{\rm N}}}}
\def\IC{\,\,{\hbox{{\rm I}\kern-.59em\hbox{\bf C}}}}
\def\IZ{{\hbox{{\rm Z}\kern-.4em\hbox{\rm Z}}}}
\def\IP{{\hbox{{\rm I}\kern-.2em\hbox{\rm P}}}}
\def\IH{{\hbox{{\rm I}\kern-.4em\hbox{\rm H}}}}
\def\ID{{\hbox{{\rm I}\kern-.2em\hbox{\rm D}}}}

\newcommand{\beq}{\begin{equation}}
\newcommand{\eeq}{\end{equation}}
\newcommand{\bea}{\begin{eqnarray}}
\newcommand{\eea}{\end{eqnarray}}

\begin{document}

\voffset 1cm

\newcommand\sect[1]{\emph{#1}---}

\title{Deconfined, Massive Quark Phase at High Density and Compact Stars: A Holographic Study}

\author{Kazem Bitaghsir Fadafan}
\affiliation{ Faculty of Physics, Shahrood University of Technology,
P.O.Box 3619995161 Shahrood, Iran}

\author{Jes\'us Cruz Rojas}
\affiliation{ STAG Research Centre \&  Physics and Astronomy, University of
Southampton, Southampton, SO17 1BJ, UK}

\author{Nick Evans}
\affiliation{ STAG Research Centre \&  Physics and Astronomy, University of
Southampton, Southampton, SO17 1BJ, UK}

\begin{abstract}
In Hoyos et al. (arXiv:1603.02943) a holographic D3/D7 system was used to describe a deconfined yet massive quark phase of QCD at finite density, concluding that the equation of state of such a phase was not stiff enough to support exotic dense stars. That analysis used a hard quark mass to represent the dynamical mass and assumed a conformal gauge background. Here we phenomenologically adjust the D3/D7 system to include a running anomalous dimension for the quark condensate. This introduces a dynamical mechanism for chiral symmetry breaking yet the model still has a deconfined massive phase at intermediate densities. We show that these systems, dependent on the running profile in the deep IR, generate much stiffer equations of state and non-monotonic behaviour in the speed of sound. The results suggest that these equations of state may be closer to supporting hybrid stars with quark cores.

\end{abstract}%

\maketitle

\newpage
\section{Introduction}

Neutron stars are unique systems where matter is at low temperatures and very high densities. The densities are high enough to consider the existence of a deconfined quark phase, but not enough to be able to apply perturbative QCD. In such compact stars it is believed that matter ranges from nuclei embedded in a sea of electrons at low densities in the crust, to the extremely neutron-rich uniform matter in the outer core, and possibly exotic states such as deconfined matter in the inner core \cite{Haensel_Neutron_Stars1}.

The equation of state (EoS) of the dense matter, which relates state variables of the system, is a key ingredient to fully model a neutron star. A complete EoS would also be very important in the light of the recent measurement of gravitational wave signals from mergers of binary neutron stars \cite{TheLIGOScientific:2017qsa}, since the model of the wave signal is sensitive to the specific form of the EoS. Nevertheless, there has been a struggle to find a complete EoS; the difficulty of the task resides in the need to solve QCD in the non-perturbative regime at finite baryon chemical potential. At the moment the EoS of  strongly interacting matter  at low temperatures is relatively well described at baryon densities below the nuclear saturation limit $n_B \leq n_s \approx 0.16 $ fm$^{-3}$, where Chiral Effective Theory (CET) works \cite{{Tews:2012fj},{Gezerlis:2013ipa}}, as well as at baryon chemical potential above $\sim 2.5$ GeV where the perturbative techniques can be applied \cite{{Freedman:1976ub},{Vuorinen:2003fs},{Kurkela:2016was}}. However this excludes the values of density where a phase transition to quark matter would be expected to occur \cite{Fukushima:2013rx}.

In the last two decades, the AdS/CFT correspondence has emerged as a new tool to study strongly coupled gauge theories \cite{Maldacena:1997re}. It provides the ability to rigorously compute in theories close to $\mathcal{N} = 4$ super Yang-Mills theory at large $N_c$ including flavour degrees of freedom \cite{Karch:2002sh, review}, using a weakly coupled gravitational dual and has provided a rich new framework for modelling other gauge systems including theories close to QCD \cite{Erlich:2005qh}. It is natural then to ask if a holographic model of the high density phase of QCD can be constructed and the corresponding EoS obtained. Holographic EoS at finite density have also been studied in \cite{{Ecker:2017fyh},{Hoyos:2016zke},{Annala:2017tqz},{Jokela:2018ers},{Li:2015uea}}.

Our goal in the present paper is to investigate whether a deconfined phase in the core of neutron stars could be stable. In \cite{Hoyos:2016zke} the authors made a first attempt at such a description using the D3/D7 system that describes quarks with a hard mass of order 330 MeV in ${\cal N}=4$ super-Yang Mills (SYM) background at finite density. Exact analytic results for the free energy are known in this case \cite{Karch:2007br}. The glue fields are deconfined, and conformal so the theory describes a putative massive, deconfined quark phase. They concluded that the equation of state was too soft to support exotic stars. However, one can critique the model since there is no chiral symmetry breaking mechanism and the hard mass is only an approximation to chiral symmetry breaking which should switch off at yet higher densities. Also since they match the conformal theory's free energy at large density to the UV of QCD they, in a sense, match the dynamics to perturbative gluons whilst one might expect a running coupling from weak to strong to have significant impact.

Here we will take a phenomenological approach to improving the D3/D7 systems predictions. We will include an effective dilaton (although it is not backreacted on the geometry) that controls by hand the running of the anomalous dimension, $\gamma$, of the quark bilinear \cite{Alvares:2012kr}. We pick a simple ansatz that has $\gamma=0$ in the UV but then runs to a dial-able fixed point value in the IR. At zero density such theories have a BKT transition as $\gamma$ in the IR is changed through one \cite{Jarvinen:2011qe, Alvares:2012kr} (the Breitenlohner Freedman bound \cite{Breitenlohner:1982jf} is violated in the model for $\gamma > 1$) from a chiral symmetric phase ($\gamma < 1$) to a chiral  symmetry broken phase ($\gamma>1$). When density is included in a theory that runs to a  fixed IR $\gamma$ we show that there are  two second order transitions - first density switches on, then at a distinct transition chiral symmetry breaking switches off. This phase structure has been seen previously in the D3/D7 system with a magnetic field \cite{{Filev:2007gb},{Evans:2010iy}} and phenomenologically related models \cite{Evans:2010hi}. Similar structures have also been seen recently \cite {Kovensky:2019bih} in the Witten Sakai Sugimoto model \cite{Sakai:2004cn}. The intermediate phase is an example of a massive yet deconfined quark phase. Our model though contains a description of a dynamical quark mass  and a running anomalous dimension. We show how the EoS in these systems depends on the UV fixed point value for $\gamma$ and show that runnings that might plausibly describe QCD have a considerable stiffer EoS than the pure D3/D7 system. The speed of sound in units of the speed of light can reach as high as $c_s^2=0.55$.

Once the EoS is obtained, solutions of the Tolman-Oppenheimer-Volkoff (TOV) equations which correspond to spherically symmetric stellar configurations that are in hydrostatic equilibrium can be found.  Nevertheless, the equilibrium of the solution does not guarantee that it is stable. There can be an instability to radial oscillations. It can be proved \cite{compactstarbook} that as one moves along the sequence of equilibrium configurations of the TOV equations, perfect fluid stars can pass from stability to instability if the equilibrium mass, $M$,  is stationary with respect to the central energy density, $\mathcal{E}_c$.  Therefore a condition for stability is that 
\begin{equation} \label{stability}
    \frac{\partial M(\mathcal{E}_c)}{\partial\mathcal{E}_c} >0.
\end{equation}
Furthermore in \cite{Alford:2017vca} the authors discuss methods for determining the stability of a star in terms of the Bardeen, Thorne and Meltzer (BTM) criteria \cite{BardeenThorneMeltzer}. 

We explore the effect of the holographic EoS we find in the TOV equation solutions. Even the stiffer possible descriptions of the deconfined quark phase we generate are not quite sufficient to construct a convincing description of both the heaviest neutron stars and new stable hybrid stars with quark matter cores. However, the situation is closer than  in the case studied in \cite{Hoyos:2016zke} and in some cases there are hints that lighter hybrid stars may exist supported by the deconfined quark matter. We report on this picture since it strongly suggests that the changes we have made are steps towards a description with interesting phenomenology and it will hopefully trigger further refinement of the holographic set up. We briefly and rather crudely discuss an example of such a refinement, adding the confinement transition as an additional shift in the pressure between the high and low density phases which may further stabilize hybrid stars although obtaining both hybrids and very heavy neutron stars remains an issue.  In future we will look to include colour superconducting phases (in the holographic spirit of \cite{BitaghsirFadafan:2018iqr}) which may further stiffen the EoS.

The paper is organized in the following way: In Section II we will review the different possible phases relevant to neutron stars - a confined phase of neutron starts which is modelled with an EoS that comes from considering a chiral effective field theory and a piecewise polytropic extension towards higher values of density; the previous work \cite{Hoyos:2016zke} implementing a deconfined phase in the neutron stars using a top-down approach to AdS/CFT and a hard mass to the quarks; and a bottom-up D3/D7 brane intersection model with a chiral symmetry breaking mechanism. In Section
III  we solve the TOV equations and analyse the mass-radius relations of neutron stars using the models of the previous section. We summarize in Section IV.

\section{The Finite Density Phase Structure of QCD}

\begin{figure}[h]
\includegraphics[width=8cm]{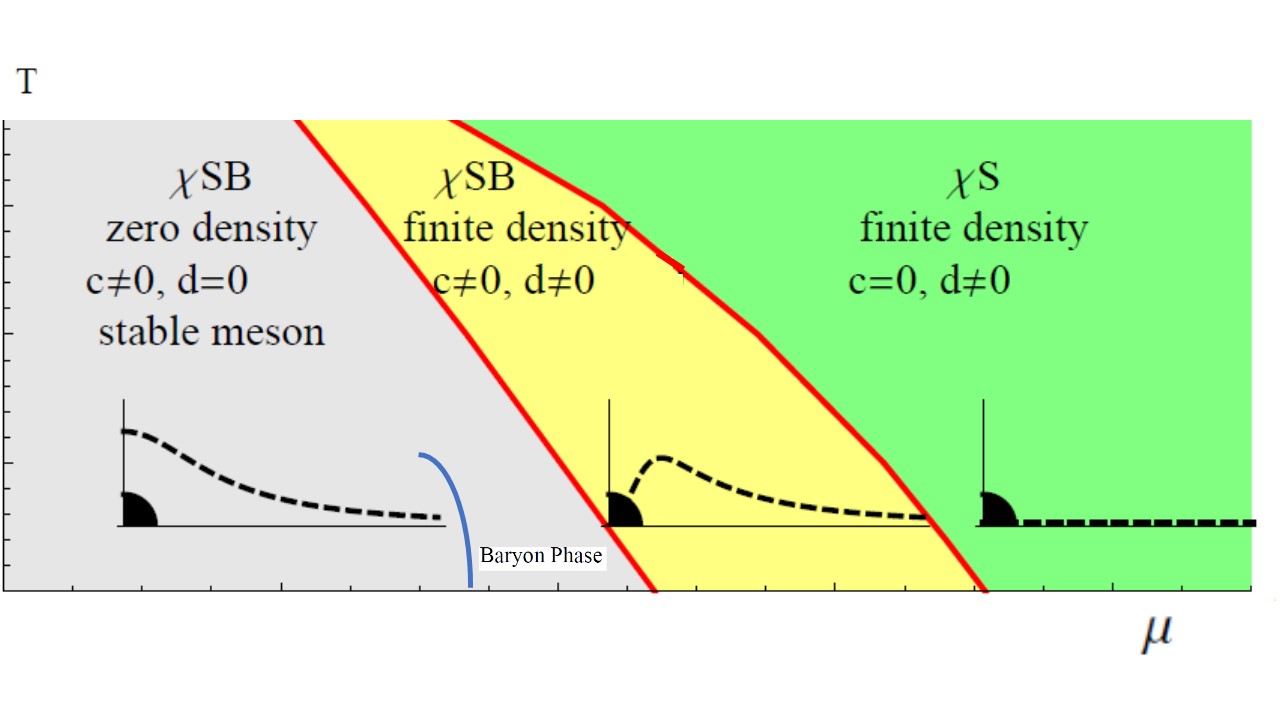}  
 
  \caption{\footnotesize{ \textit{ A sketch of the low temperature phase structures we observe in the holographic models we explore. At low chemical potential the theory has chiral symmetry breaking ($\chi$SB), a non-zero chiral condensate ($c$) and zero density ($d$); in an intermediate regime there is a deconfined massive quark phase with non-zero density; at high $\mu$ there is chiral symmetry restoration. The D7 embedding function (field $\chi$) is also sketched in each phase. These transitions are all second or higher order in the holographic models. Note we have also sketched the position of the baryon phase with non-zero neutron density which is not present in the holographic models (we include it phenomenologically from low energy analysis) - we expect the transition to the high density phases from the baryon phase to be first order. }}}
  \label{fig:1}
\end{figure}

In this section we will review our model of the low temperature QCD phase structure and the models that we use to study each phase. In Figure \ref{fig:1} we sketch the phase structures that we will see below as a function of quark chemical potential at low T. In fact in this paper we will only compute at strictly T=0 although holography would straightforwardly allow computation at finite T also. 

\subsection{Nuclear phase}
At small chemical potentials QCD is well understood. The confined, chirally broken vacuum is empty until a chemical potential of $\mu = 308.55$ MeV when there is a first order phase transition to nuclear matter. This transition is already well studied and the nuclear matter equation of state has been explored in \cite{Hebeler:2013nza}. There the authors combined observations of a $1.97$ solar mass neutron star with effective field theory (EFT), thereafter extrapolating it with a constrained piecewise polytropic form. Here holography is probably least able to help - given its origin at infinite $N_c$ baryons are naturally very heavy and far from the QCD limit so, following several other authors \cite{{Hoyos:2016zke},{Ishii:2019gta},{Jokela:2018ers}}, we will simply use  the results of \cite{Hebeler:2013nza} to model the nuclear phase. Note there have been attempts to study the QCD nuclear phase holographically, for example in \cite{Bergman:2007wp, Li:2015uea,Evans:2012cx}, but this will not be our focus in this paper.

\begin{figure}[h]  
 \includegraphics[width=8cm]{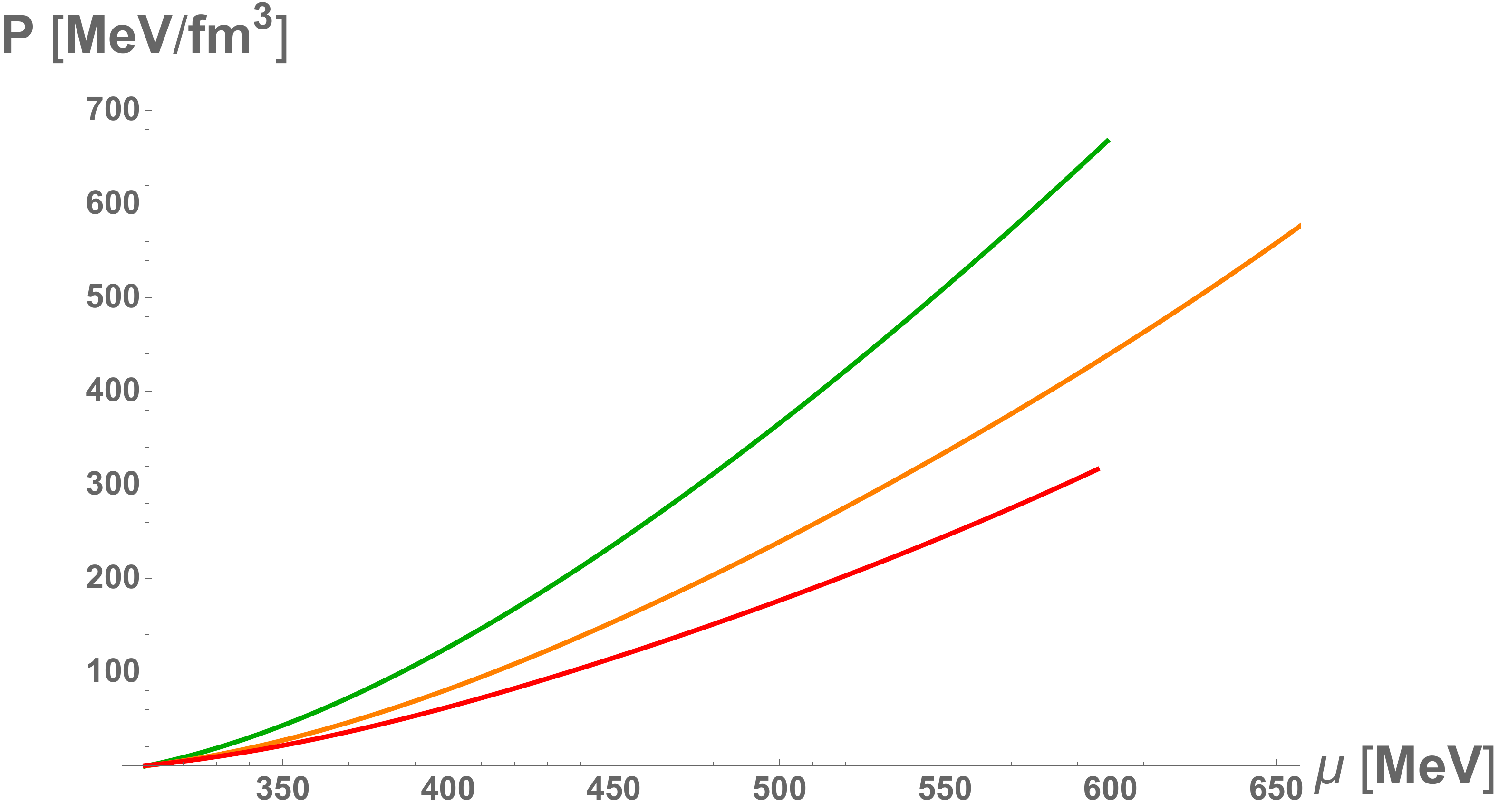}  
 \includegraphics[width=8cm]{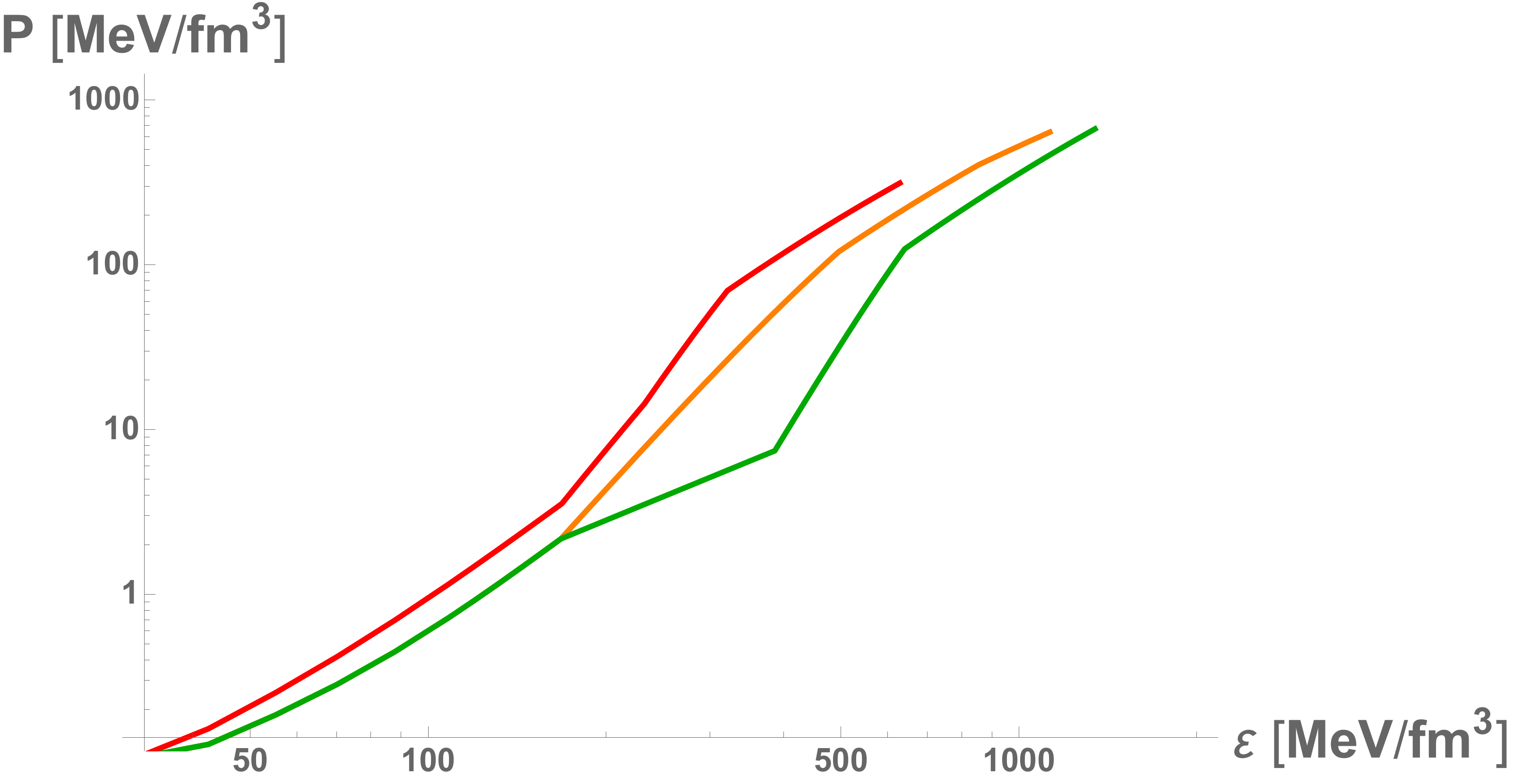}
  \caption{\footnotesize{ \textit{Data for the nuclear phase taken from  \cite{Hebeler:2013nza}: we show both the pressure versus chemical potential and energy density.  The Green line represents a soft EoS, the orange a medium EoS and the red line a stiff EoS.}}}
  \label{fig:2}
\end{figure}

\begin{figure}[h]
 \includegraphics[width=8cm]{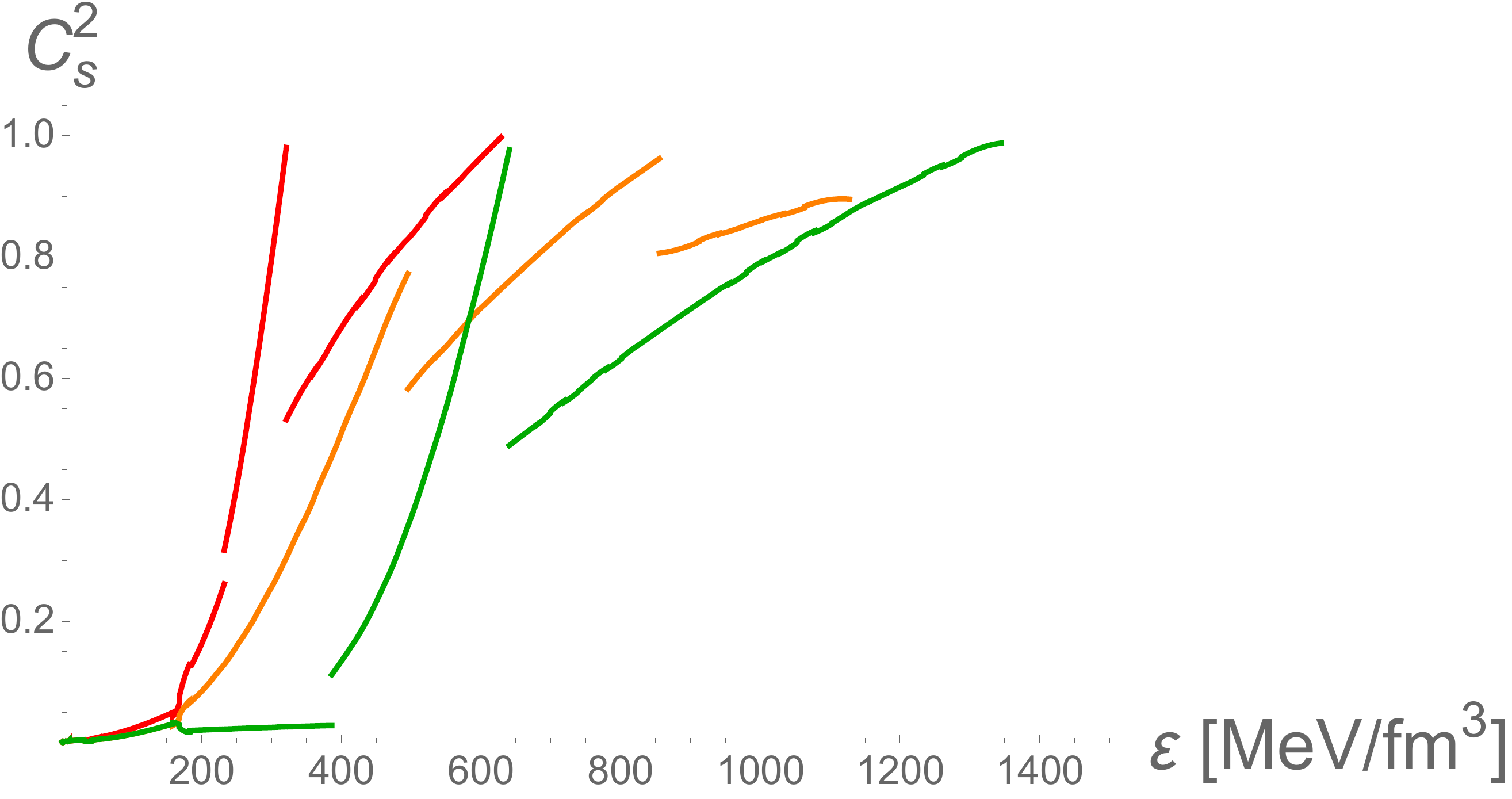}  
  \caption{\footnotesize{ \textit{Speed of Sound squared as a function of the energy density for nuclear matter \cite{Hebeler:2013nza}. The different coloured lines represent nuclear matter from EFT EoS; (Green) soft EoS, (orange) medium EoS and (red) stiff EoS.}}}
  \label{fig:3}
\end{figure}

Three ansatz for the EoS (soft, medium and stiff) are presented in Table 5 of \cite{Hebeler:2013nza} - they give the energy density and pressure for different densities. A stiff equation of state is one where the pressure increases quickly for a given increase in density. Such a material would be harder to compress and offers more support against gravity. Conversely, a soft equation of state produces a smaller increase of pressure for a change in density and is easy to compress. We have encoded their data as a Mathematica fitting polynomial for the analysis below and we plot these in Figure \ref{fig:2}. We will see (Figure \ref{fig:9}) that this is sufficient to reproduce the neutron star mass radius plots in \cite{Hebeler:2013nza}.

For each EoS there is a maximum central pressure/energy density for which data is provided in \cite{Hebeler:2013nza}. Above this maximum pressure either the speed of sound (which is simply the square root of ${\partial P \over \partial \mathcal{E}}$) grows greater than the speed of light or the EoS has been sufficient to explain the most massive neutron stars observed. In figure \ref{fig:3} we plot the square of the speed of sound against 
$\mathcal{E}$ to show this behaviour (note the discontinuities reflect moves between different polytropes in the piecewise construction of the equation of state in \cite{Hebeler:2013nza}) - the equivalent maximum pressures for the three possible EoS are 312.6 MeV fm$^{-3}$ (stiff)  637.2 MeV fm$^{-3}$ (medium) 666.5 MeV fm$^{-3}$ (soft).

\subsection{Holography of a Deconfined Massive Quark Phase}

The next expected transition beyond the nuclear phase as the chemical potential is raised is normally presented as a transition to a deconfined, chirally symmetric quark phase. The transition from the nuclear matter phase is normally assumed to be first order  although since this regime lies outside the region of controlled computation this is fundamentally a guess. 

Holography can potentially inform us about the transition from the empty low $\mu$ vacuum to the higher $\mu$ vacuum with non-zero quark density. The first paper studying neutron stars using holographic equations of state was \cite{Hoyos:2016zke}. There, the authors used the equation of state of the massive D3/D7 system at finite density \cite{Karch:2007br} to describe the quark matter phase. The D3/D7 model at finite density is always deconfined in the large $N_c$ limit and further has no chiral symmetry breaking mechanism. This phase naively therefore has deconfined massless quarks.  The authors then included a bare (hard) quark mass of order $\Lambda_{QCD}$ as an approximation to a chirally broken state. This is a simplistic approximation to a phase of deconfined yet massive quarks. Inherently there is an assumption here that confinement and chiral symmetry breaking transitions are separated in the high density phase structure and we will further consider such a possibility in this paper.

There is evidence for such a phase in more refined D3/D7 systems with explicit chiral symmetry breaking dynamics (see \cite{review} for examples of adding chiral symmetry breaking to the D3/D7 system). The most controlled case is where a magnetic field is introduced \cite{Filev:2007gb} - the phase diagram was generated in \cite{Evans:2010iy}. It has the structure shown in Figure \ref{fig:1} - there is a low $\mu$ phase with chiral symmetry breaking and no density. A  second order  transition then takes the model to a phase with non-zero density but chiral symmetry breaking which is precisely such a massive deconfined phase. Then another second order transition moves the system to a dense but chirally symmetric phase. Other examples of these transitions have been explored in \cite{Evans:2010hi}. The phenomenological model we use below is motivated by this example but allows one to control the running of the quark bilinear anomalous dimension $\gamma$ by hand. The key role of this running for chiral symmetry breaking was highlighted in \cite{Jarvinen:2011qe} and adapted to the D3/D7 system in \cite{Alvares:2012kr}. Our model has the advantages of an explicit chiral symmetry breaking mechanism, a running $\gamma$ and a very high $\mu$ phase with chirally symmetric quarks.  Note though none of these models naively include confinement of the gluon degrees of freedom - we will discuss this issue more in section IIIC.

In this subsection we will review the original D3/D7 model and then provide a more sophisticated D3/D7 inspired phenomenological model that has a chiral symmetry breaking mechanism built in and naturally generates this massive deconfined phase. 

\subsubsection{The Basic D3/D7 Model} 

Let us quickly review the model of \cite{Hoyos:2016zke}. Their base model is $\mathcal{N}=2$ SYM with the matter content of $\mathcal{N}=4$ $SU(N_c)$ SYM in the adjoint sector and $N_f$ matter hypermultiplet in the fundamental representation.  The DBI action for a probe D7 brane in pure AdS, with a constant dilaton, is
\begin{equation}
S = - {N_f N_c \over \lambda} T_{D7} V_3 \int d \rho \rho^3 \sqrt{1 + (\partial_\rho \chi)^2 - 2 \pi \alpha' (\partial_\rho A_t)^2} \end{equation}
Here  $\lambda$ is the `tHooft coupling, $T_{D7} = (2 \pi)^{-7} \alpha^{\prime -4}$ is the D7 brane tension, $V_3= 2 \pi^2$ is the volume of the $S^3$ on the D7 brane and $\rho$ the radial direction in AdS$_5$. $\chi(\rho)$, the brane embedding function, is holographically dual to the quark mass and condensate and $A_t$ is a gauge field dual to the quark number chemical potential and density.    In practice one works with  the action
\begin{equation}
S = - \int d \rho ~ \rho^3 \sqrt{1 + (\partial_\rho \chi)^2 -  (\partial_\rho A_t)^2} \label{simple} \end{equation}
then an analytic form for the free energy can be found \cite{Karch:2007br}
\begin{equation}  \label{HoyosF} {\cal F} = {1 \over \eta^3} (\mu^2 - m^2)^2  + {\cal O}(\mu^3T, T^4) \hspace{0.5cm} \eta = {\Gamma (7/6) \Gamma (1/3) \over \sqrt{\pi}} \end{equation}
where $m$ and $\mu$ are the UV asymptotic values of $\chi$ and $A_t$ respectively. The field theory mass and chemical potential are given by $(2 \pi \alpha^\prime)^{-1}$ times these quantities. Note that $\alpha^\prime$ (which is formally zero in the supergravity limit) then cancels from the resulting free energy for the field theory, as usual in the AdS/CFT correspondence.

To match the asymptotic UV form known from QCD one can pick $\lambda= 3 \pi/\eta^3$ so that:
\begin{equation}  {\cal F} =  {N_c N_f \over 12 \pi^2} \mu^4. \label{uv} \end{equation} 
In practice one computes with (\ref{simple}) and rescales by $ {N_c N_f  \eta^3\over 12 \pi^2}$. We will use $N_f=N_c=3$.  

Note that at any non-zero T this theory is deconfined. The phase therefore describes a vacuum with a density of quarks of mass $m$.

The EoS, which relates the pressure $P$ to the energy density $\mathcal{E}$ is found from
\begin{equation}
\begin{aligned}
P=-\mathcal{F}, \qquad \mathcal{E}=\mu\frac{\partial P}{\partial \mu}-P.
\end{aligned}
\end{equation}

The authors of \cite{Hoyos:2016zke} match this quark matter description with the nuclear EoS from the previous section to model a transition between confined and deconfined matter inside a neutron star. They equated the zero $\mu$ phases in the nuclear model of QCD and in the D3/D7 system. This  allows comparison of the nuclear phase's free energy, with the free energy of the holographic model at finite $\mu$ and then determines the dominant phase at each quark chemical potential. The hard mass of the quarks is a free parameter and, as can be seen from (\ref{HoyosF}), the phase transition occurs at $\mu=m$ when the free energy rises from zero (the phase with density does not exist for $\mu < m$). 

\begin{figure}[h] 
 \includegraphics[width=8.5cm]{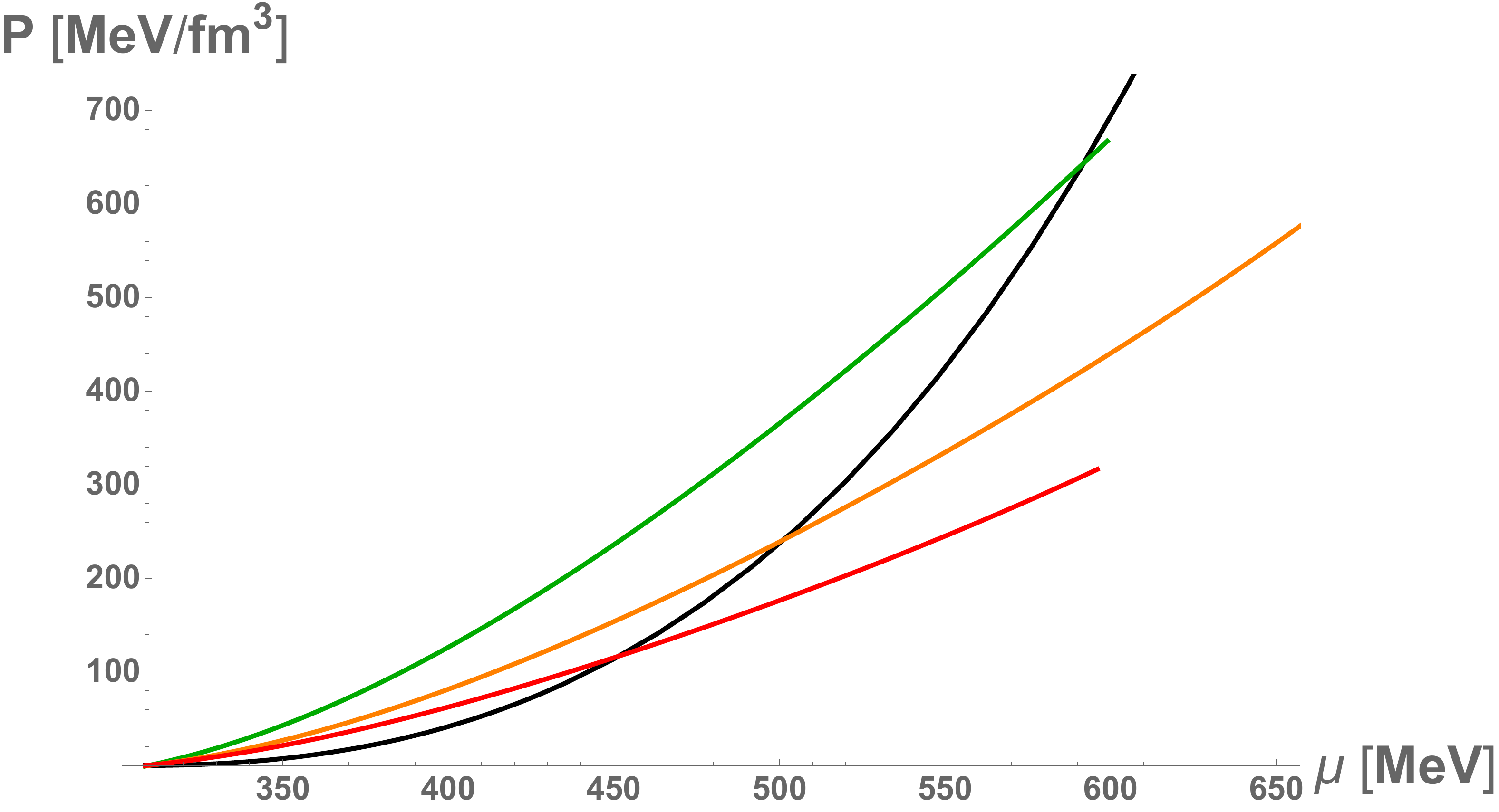}  
 \includegraphics[width=8.5cm]{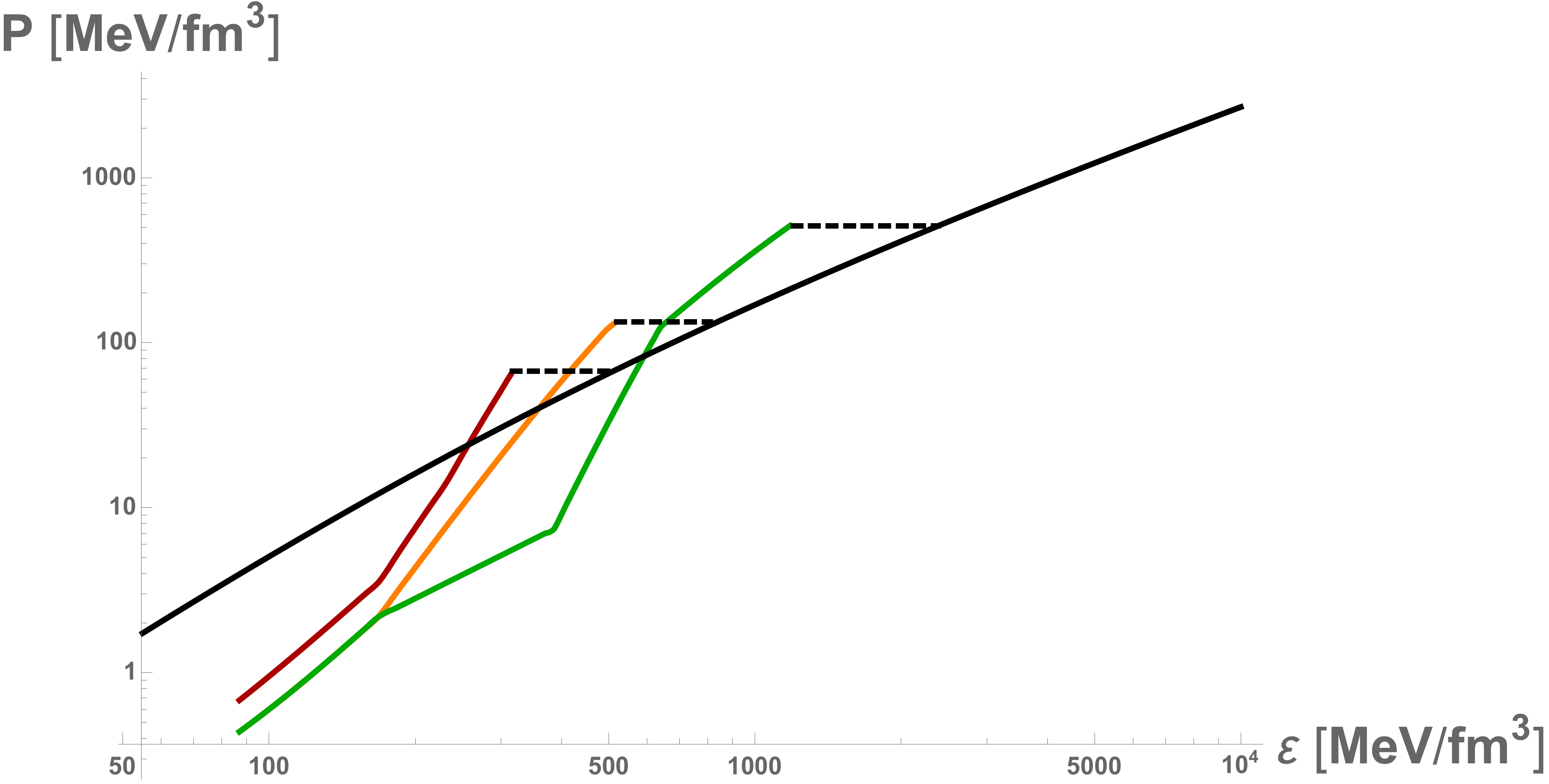} 
  \caption{\footnotesize{ \textit{ Pressure versus $\mu$ and energy density for the basic D3/D7 model of \cite{Hoyos:2016zke} in solid black. Coloured lines represent the nuclear matter from the EFT EoS. The horizontal black dotted lines shows the change of phase.}}}
  \label{fig:4}
\end{figure}

In \cite{Hoyos:2016zke} the authors set, somewhat arbitrarily, $m=308.55$MeV which places the transition to  the nuclear phase in one model and that to the deconfined massive quark phase in the other at the same critical $\mu$. We reproduce the plots for this case in Figure \ref{fig:4}. The transition between the nuclear and  deconfined massive phases occurs at the value of $\mu$ where the pressure of the deconfined quarks is greater than the chosen nuclear phase. The nuclear phase is preferred at $\mu$ just above 308.55 MeV but then there is a transition to the deconfined massive phase (note in each case before the nuclear phase reaches the pressure at which the speed of sounds becomes too large). We also display the pressure versus energy density plot which shows a jump at the first order transition.

In the later paper \cite{Annala:2017tqz} the authors allowed the critical $\mu$ of the massive deconfined phase to vary by simply dialling the quark mass $m$. If it is pushed higher than  308.55MeV the transitions occur at higher $\mu$. The authors also proposed moving the critical $\mu$ less than 308.55MeV. Now the massive deconfined phase is favoured at $\mu$ less than 308.55MeV but they showed that in intermediate regions the nuclear phase could be favoured. This leads  to compact stars with a variety of quark and neutron layers. This is quite a radical view of the phase structure although not obviously impossible. We will not consider such cases further here though. Here we will always assume any quark phase lies at $\mu$ above where the nuclear phase exists.

\subsubsection{Bottom-Up D3/D7 model with chiral symmetry breaking mechanism}

The first new question we wish to ask is how robust the simple D3/D7 model's predictions are? In particular it is a very rough and ready description of a massive deconfined quark phase with chiral symmetry breaking since the quark mass is put in by hand as a hard mass. In particular since the gauge coupling of ${\cal N}=4$ SYM is conformal one would expect the IR action to not reflect the growth of the gauge coupling. It is quite simple to construct a D3/D7 inspired bottom-up model with an explicit chiral symmetry breaking mechanism that realizes the deconfined yet massive quark phase. Here we will follow this path to cross check the results with those of the simpler model. 

Our simple model is a small variation of the DBI action for a probe D7 brane in AdS$_5$ 
\begin{equation}  \label{action}
    \mathcal{L}=-\int d\rho ~h[\rho^2 + \chi^2] ~\,\rho^3 \sqrt{1+(\partial_{\rho}\chi)^2-(\partial_{\rho}A_t)^2} 
\end{equation}
The function $h$ which is crucially a function of $(\rho^2 + \chi^2)$ is the key extra ingredient beyond (\ref{simple})- an effective dilaton term. In top down models the dilaton will be constant for ${\cal N}=4$ SYM or for more complicated cases backreact on the metric. Here in a bottom-up approach we will allow $h$ to be non-trivial yet neglect any backreaction in the metric. $h$ will trigger chiral symmetry breaking. Note an explicit top-down example of precisely this action and a non-trivial, yet not backreacted, $h[\rho^2 + \chi^2]$ that causes breaking of the symmetry is obtained for the example of magnetic field $B$ induced chiral symmetry breaking in \cite{Filev:2007gb}.  We will restrict ourselves to functions for $h$ that return in the UV to a constant so that the  UV normalization follows that of the discussion around (\ref{uv}).

Naively one might think to use the running coupling from QCD as the ansatz for the dilaton $h$. However, in \cite{Jarvinen:2011qe, Alvares:2012kr} it was shown that the mapping of the dilaton to the running anomalous dimension of the $\bar{q}q$ operator, that determines the chiral symmetry breaking dynamics, is more subtle. In particular chiral symmetry breaking is triggered when the chirally symmetric embedding $\chi=0$ becomes unstable. One can expand the action for small $\chi$ \cite{Alvares:2012kr} to give
\begin{equation}
S \simeq \int d \rho ~ \left[{1 \over 2} h|_{\chi=0} ~ \rho^3 (\partial_\rho \chi)^2 + \rho^3 \left. {\partial h \over \partial \chi^2} \right|_{\chi=0} \chi^2 \right] \label{expand}
\end{equation}
The first term can be made the kinetic term of a canonical scale in AdS$_5$ by writing $\chi= \tilde{\rho} \phi$ with the coordinate change
\begin{equation}
\tilde{\rho} =    {1 \over \sqrt{2}}     \left( \int_\rho^\infty h^{-1} \rho^{-3} d \rho \right)^{-1/2}
\end{equation}
leaving
\begin{equation} S \simeq \int d \tilde{\rho} {1 \over 2} \left( \tilde{\rho}^5 (\partial_{\tilde{\rho}} \phi)^2 - m^2 \phi^2 \right) 
\end{equation}
with 
\begin{equation} \label{runmass}
m^2 = -3 + h {\rho^5 \over \tilde{\rho}^4} {d h \over d \rho}
\end{equation}
As expected the field $\chi$ maps to a field $\phi$ with $m^2=-3$ in the case where $h=$ constant - it holographically describes the mass and quark condensate of dimensions 1 and 3 (satisfying the required $m^2 = \Delta (\Delta-4)$). When $h$ is $\rho$ dependent in the IR though there is an additional contribution to $m^2$, a running of $\Delta$. If $m^2$ passes through $-4$ then the Breitenlohner Freedman (BF) bound in AdS$_5$ is violated, there is an instability and the D7 embedding function moves away from $\chi=0$ - chiral symmetry is then broken.

Thus $h=$ constant describes a theory with no anomalous dimension. In \cite{Alvares:2012kr} it was shown that $h=1/\rho^q$ describes a theory with 
\begin{equation} m^2 = -3 - \delta m^2, ~~~~ \delta m^2 = {4 q \over (2-q)^2}
\end{equation}
$m^2=-4$ is achieved when $q=0.536$ and it becomes infinite at $q=2$. In terms of the anomalous dimension of the IR phase we have
\begin{equation} \gamma = 1 - \sqrt{1- {4 q \over (2 -q)^2}}. \end{equation}

It's worth stressing that this analysis in a sense legitimises not backreacting the dilaton factor in our model. If one did have a fully backreacted geometry then the expansion to (\ref{expand}) would be more complicated but the additional pieces from expanding metric terms and so forth would simply be an additional contribution to the running mass in (\ref{runmass}). At the level of studying the instability to chiral symmetry breaking putting in a hand chosen dilaton is as good as including a more elaborate bottom up geometry (of course if one had an honest full description of the particular chiral symmetry breaking system then the subtleties would be important!). 

A natural choice to describe the running in a QCD like theory is
\begin{equation} \label{hform}
h = 1 + {1 \over (\rho^2 + \chi^2)^{q\over2}}
\end{equation}
which has zero anomalous dimension in the UV whilst moving to an IR regime below $(\rho^2 + \chi^2)^{1\over2}=1$ with a fixed point for the anomalous dimension.  Note we include $\chi$ here in the spirit of the D3/D7 models we have discussed. Importantly if it were not present the BF bound would be violated in the model no matter how large $\chi$ became so there would be no stable solutions for $\chi$.

There is intrinsically a single scale in this ansatz (the numerator of the fraction), which we have set to 1, and it loosely sets units where $\Lambda_{QCD}$=1.
In fact this scale represents where the model moves from weak coupling to strong coupling. For the walking theories this scale may be quite separated from the IR scale where the BF bound is violated and chiral symmetry breaking occurs. We find it more intuitive therefore below to write all physical observables in units of the IR quark mass  $\chi_0= \chi(0)$ which for comparison to QCD should be taken to be 330 MeV or so. There is still though only the single scale in the model.

By varying $q$ one can pick very walking theories \cite{Holdom:1981rm} where the anomalous dimension asymptotes to the BF bound at $q=0.536$ or theories that run quickly to large IR fixed points $q \simeq 2$. There are also theories that have a divergent anomalous dimension at some finite value of $(\rho^2 + \chi^2)^{1\over2}$ by picking $q>2$. It is interesting in this latter case that the anomalous dimension diverges at some finite energy scale (as it would at one or two loop level in QCD) yet the gravity dual provides a smooth description below that scale. It is a matter of speculation as to the IR behaviour of the QCD running and we will explore a range of possible IR divergent and fixed point behaviours below. The theory is known not to be very walking though so values of $q$ towards 2 are most likely appropriate.  In \cite{Alvares:2012kr} it was shown that the zero density chiral transition as one varies $q$ shows BKT or Miransky scaling \cite{Miransky:1996pd,Kaplan:2009kr} because the IR mass is smoothly tuned through the BF bound. 

The reader might wonder how generic our ansatz for $h$ is.  In the UV it must be a constant (so $\gamma=0$). In the IR it must go as $1/\rho^q$ at $\chi=0$ (so $\gamma$ has the desired fixed point value). Writing the AdS radial direction as $(\rho^2 + \chi^2)^{1\over2}$ is correct in the probe D7 model and dimensionally correct - it seems sensible. The question then is about the transition region between the IR and UV regimes - here we have picked something generically monotonic. In fact in \cite{Evans:2013vca} we argued that in these chiral symmetry breaking models the physics is determined by the derivative of gamma at the scale of the BF bound violation. Here our function simply raises this derivative as $q$ and the fixed point value of $\gamma$ rises. Of course, one could imagine doing some wilder things where there are, for example, multiple plateaus in the running but given nothing like that is well motivated we believe our ansatz is in fact reasonably generic.

Our theory then is (\ref{action}) with (\ref{hform}). Note that in the large $\rho$ limit these theories return to the description of  \cite{Hoyos:2016zke} since $h \rightarrow 1$ so we fix the coefficient of the Lagrangian as in \cite{Hoyos:2016zke} to match to the asymptotic perturbative prediction of the free energy from QCD - that is we enforce (\ref{uv}) in the UV.

Since the Lagrangian does not depend on the field $A_t$ we have a conserved constant, the density, $d=\frac{\delta \mathcal{L}}{\delta A'_t}$, from here we can find an equation for $A_t$. Then we can perform a Legendre transformation $\mathcal{L}'=\mathcal{L}-A'_t \frac{\delta \mathcal{L}}{\delta A'_t}$ to replace $A_t$ by $d$ in the Lagrangian and find an equation for $\chi$. The equations of motion are
\begin{equation} \label{mueqn}
(\partial_{\rho}A_t)^2=\frac{d^2(1+(\partial_{\rho}\chi)^2)}{h^2 \rho^6+d^2}, 
\end{equation}

\begin{equation} \label{Leqn}  \begin{array}{l}
\partial_{\rho}\left(\frac{\left(h^2\rho^6+d^2\right)\partial_{\rho}\chi}{\sqrt{(1+(\partial_{\rho}\chi)^2)(h^2\rho^6+d^2)}}  \right)\\
\\
-\frac{(1+(\partial_{\rho}\chi)^2)\rho^6 h\frac{\partial h}{\partial \chi}}{\sqrt{(1+(\partial_{\rho}\chi)^2)(h^2\rho^6+d^2)}} = 0 \end{array}
\end{equation}
which we then numerically solve. 

\begin{figure}[]
 \includegraphics[width=7.5cm]{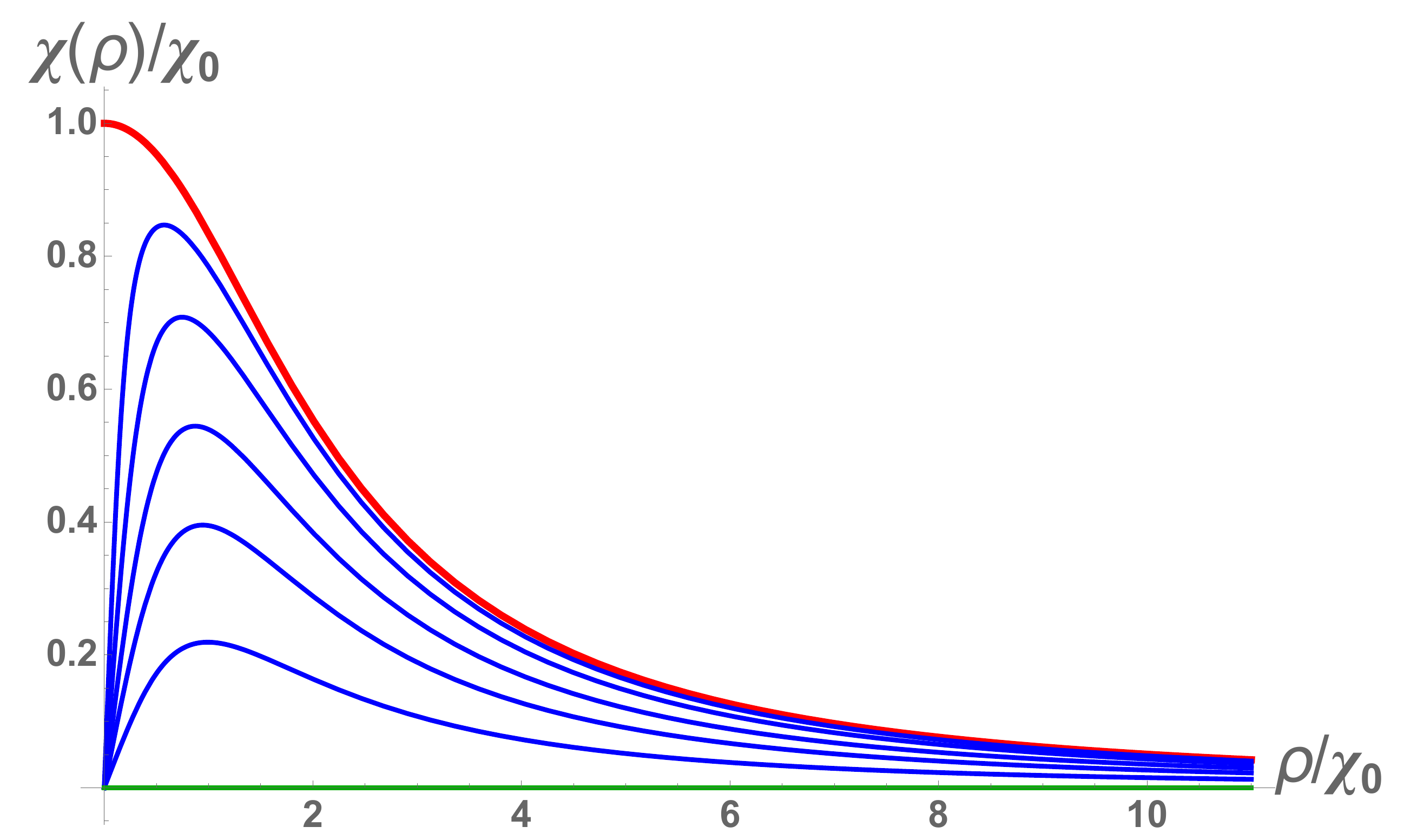} 
  \caption{\footnotesize{ \textit{Solutions for $\chi(\rho)$ for the case $q=1.8$ in equation (\ref{hform})  for $d=0$ (Red), $d=0.005, 0.015, 0.075, 0.15, 0.29$ (from top to bottom in Blue) and $d=0.501$ (Green).}}}
            \label{fig:5}
\end{figure}

First consider the case where $d=0$, the low chemical potential phase, we fix the initial condition $\chi'(0)=0$ and tune $\chi(0)=\chi_0$ (these are the standard IR boundary conditions in such models) in order that the UV mass obtained from the large $\rho$ behaviour of $\chi(\rho)$ is zero. We display the solution in red in Figure \ref{fig:5} for the case $q=1.8$: the function $\chi(\rho)$ can be viewed as the dynamical mass function of the quarks - in the UV (large $\rho$) limit the bare mass is zero, but as one runs to the IR (low $\rho$) a dynamical mass switches on.

\begin{figure}[]
 \includegraphics[width=7.5cm]{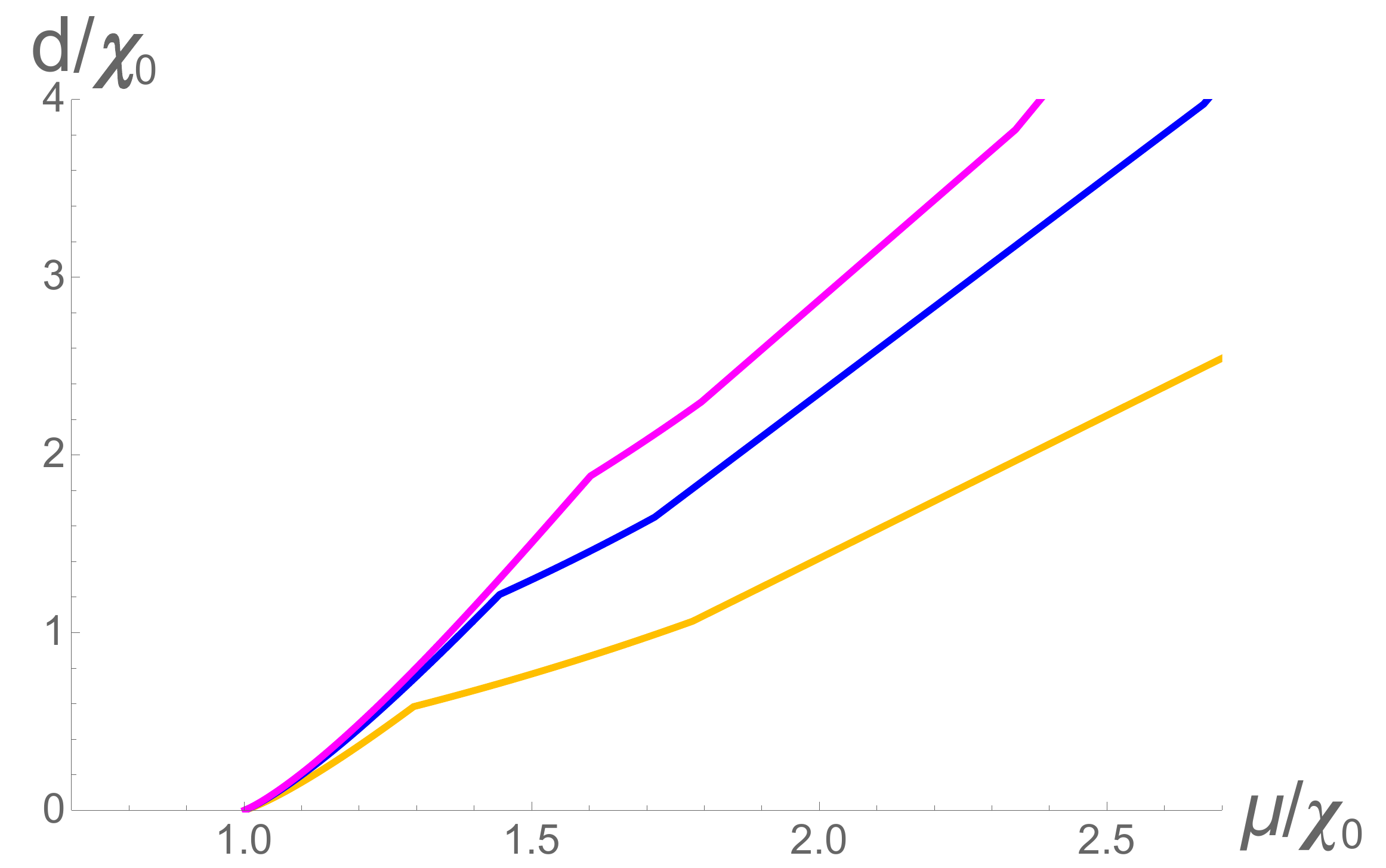} 
  \caption{\footnotesize{ \textit{d vs $\mu$ for different values of $q$. The different coloured lines represent different values of q; (yellow) q=1.6, (dark blue) q=1.99,  (magenta) q=2.3}}}
            \label{fig:6}
\end{figure}

In the large chemical potential phase we vary the value of $d$ which is in correspondence to the chemical potential through (\ref{mueqn}). We set $A_t(0)=\chi(0)=0$ and vary $\chi'(0)$ (again standard D3/D7 boundary conditions with density \cite{Kobayashi:2006sb}) for each value of $d$ in order to obtain solutions that have a UV mass equal to zero - see the blue curves in Fig \ref{fig:5} in the case of $q=1.8$. We also obtain the value of the chemical potential as the UV value of $A_t$, i.e $\mu=A_t(\rho \rightarrow \infty )$ from integrating (\ref{mueqn}). We find that there is a critical value $d_c$ above which  there is not a symmetry breaking process and then the only solutions with a zero UV mass are the solutions that have $\chi=0$ for every value of $\rho$ (green in Fig \ref{fig:5}). There are two second order transitions here, from the red $d=0$ solution to the blue chiral symmetry breaking solutions, which is the massive deconfined phase we discuss, to the green very large $d$ chirally symmetric phase.  In Figure \ref{fig:6} we show the density as a function of $\mu$ for several different $q$. It displays the two transitions - one where $d$ switches on and the second is where there are kinks, corresponding to the point where the condensate switches off. At each transition $d$ is continuous but there is a discontinuity in the derivative showing the transitions are second order. 

We obtain the free energy of the vacuum for each value of $d$
by integrating the action using the solutions of (\ref{Leqn}). The integrals all share the same divergence which can be removed by subtracting the counter term $\int d\rho \rho^3$. We further subtract the $d=0$ free energy from the $d\neq0$ solutions free energies so that the vacuum at low $\mu$ has ${\cal F} =0$ as assumed in the previous nuclear equation of state analysis. Since $d$ is related to $\mu$ we can obtain results as a function of the chemical potential. 

\begin{figure}[]
 \includegraphics[width=7.5cm]{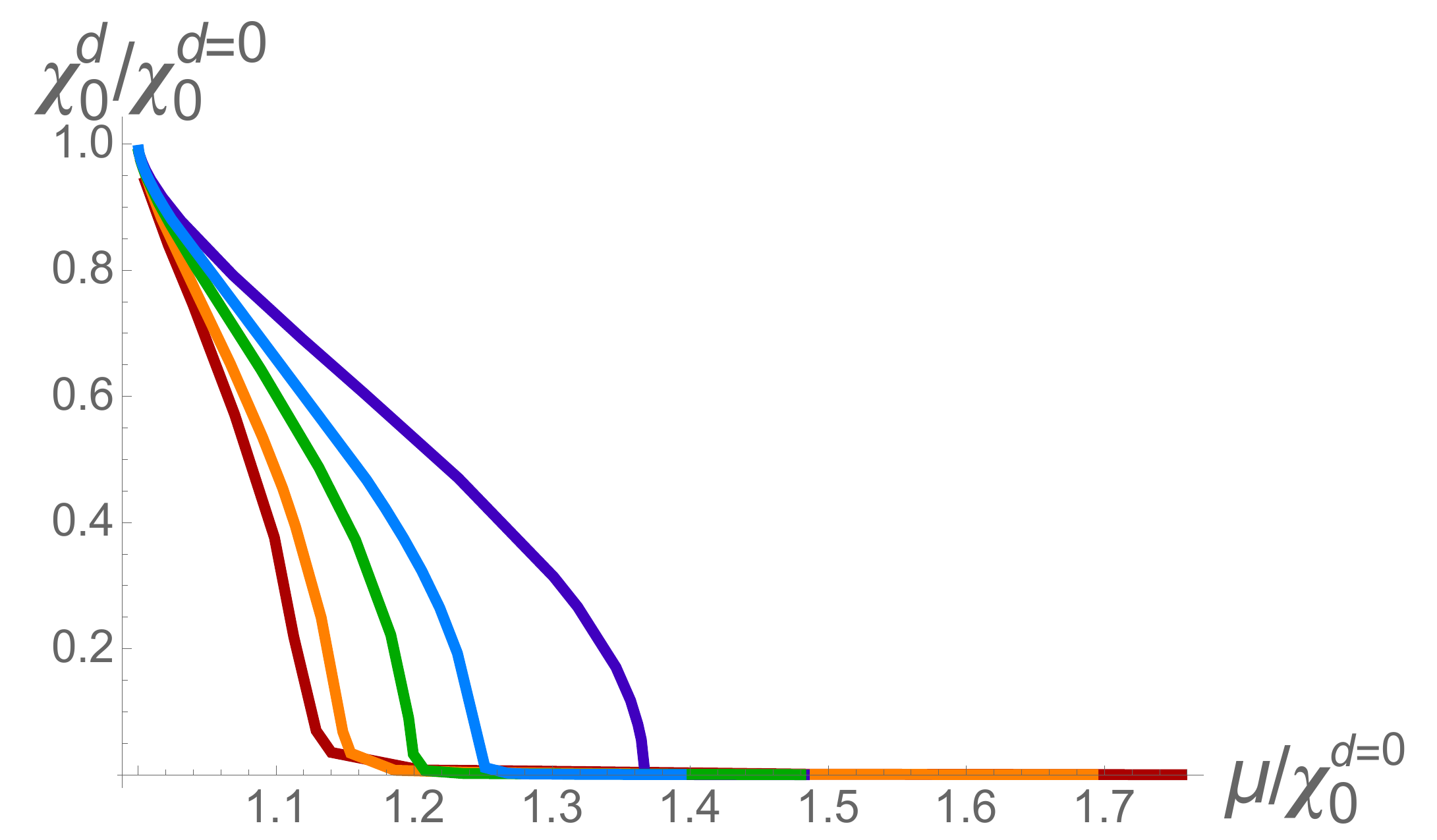} 
  \caption{\footnotesize{ \textit{$\chi_{max}$ vs $\mu$ for different $q$. The different coloured lines represent different values of q; (Red) q=1, (orange) q=1.1,  (green) q=1.3, (blue) q=1.45, (purple) q=1.8.}}}
            \label{fig:65}
\end{figure}

Now we can study the behaviour of the model as a function of $q$. To make this comparison fair we write all dimensionful parameters in units of $\chi_0=\chi(0)$ at $\mu=0$ - this can be thought of as the constituent quark mass (naively $\simeq 330 $ MeV, a third the proton mass) which we are then using to fix the comparison. First of all we can look at the phase structure with chemical potential - in Fig \ref{fig:65} we display the peak value of the embedding $\chi(\rho)$  against $\mu$ for different $q$. The larger $q$ values represent high IR fixed point theories with strong running as the BF bound is violated and they more strongly support the embedding $\chi$ as $\mu$ rises but then rather rapidly switch to the $\chi=0$ phase. Lower $q$ theories that have smaller IR fixed point values support the peak of $\chi(\rho)$ less well but the chirally broken phase persists to higher $\mu$ - this supports the idea that the $\chi(\rho)$ functions have support in the more walking theories to higher energy scales.  

\begin{figure}[]
 \includegraphics[width=7.5cm]{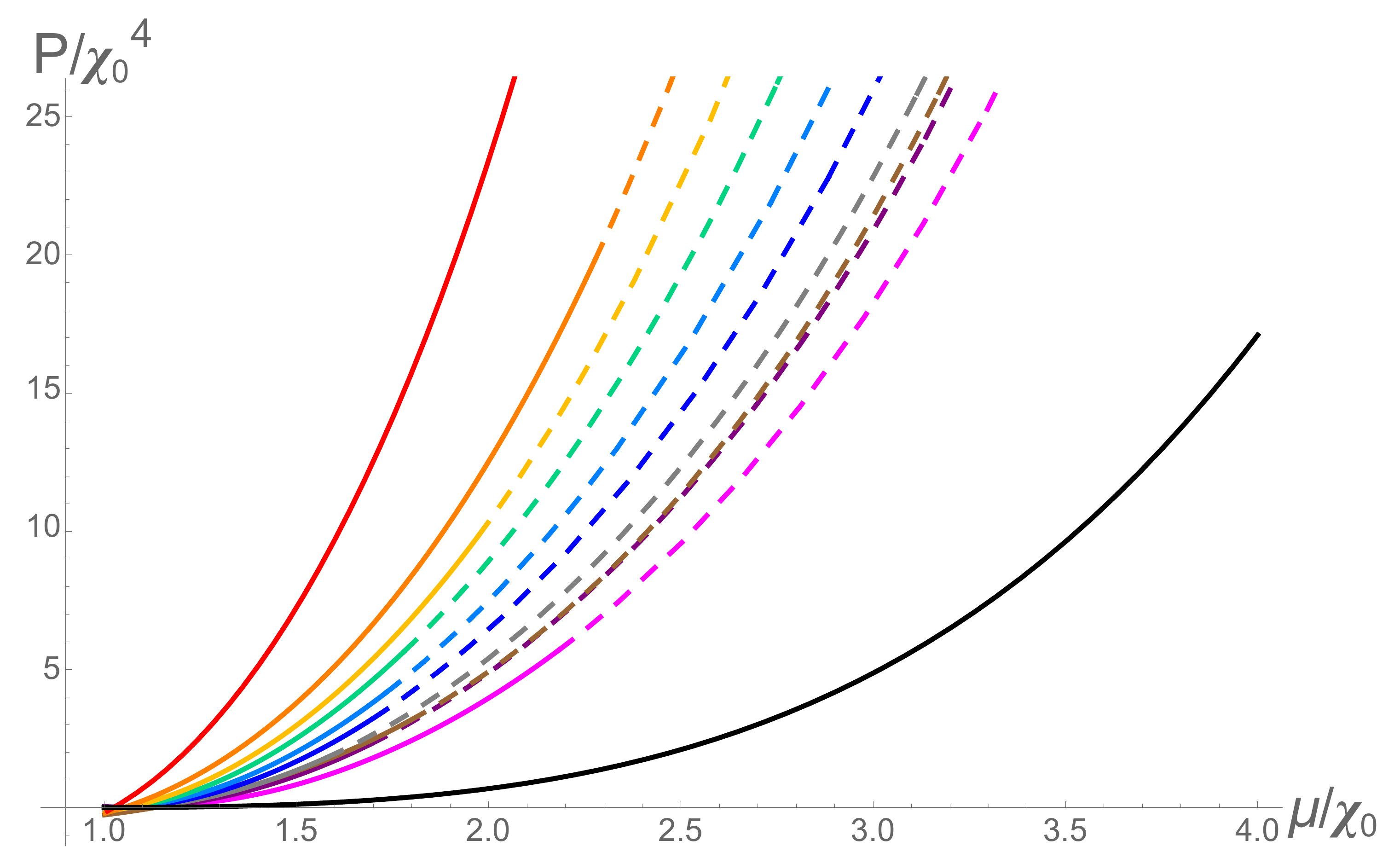}\\
 \includegraphics[width=7.5cm]{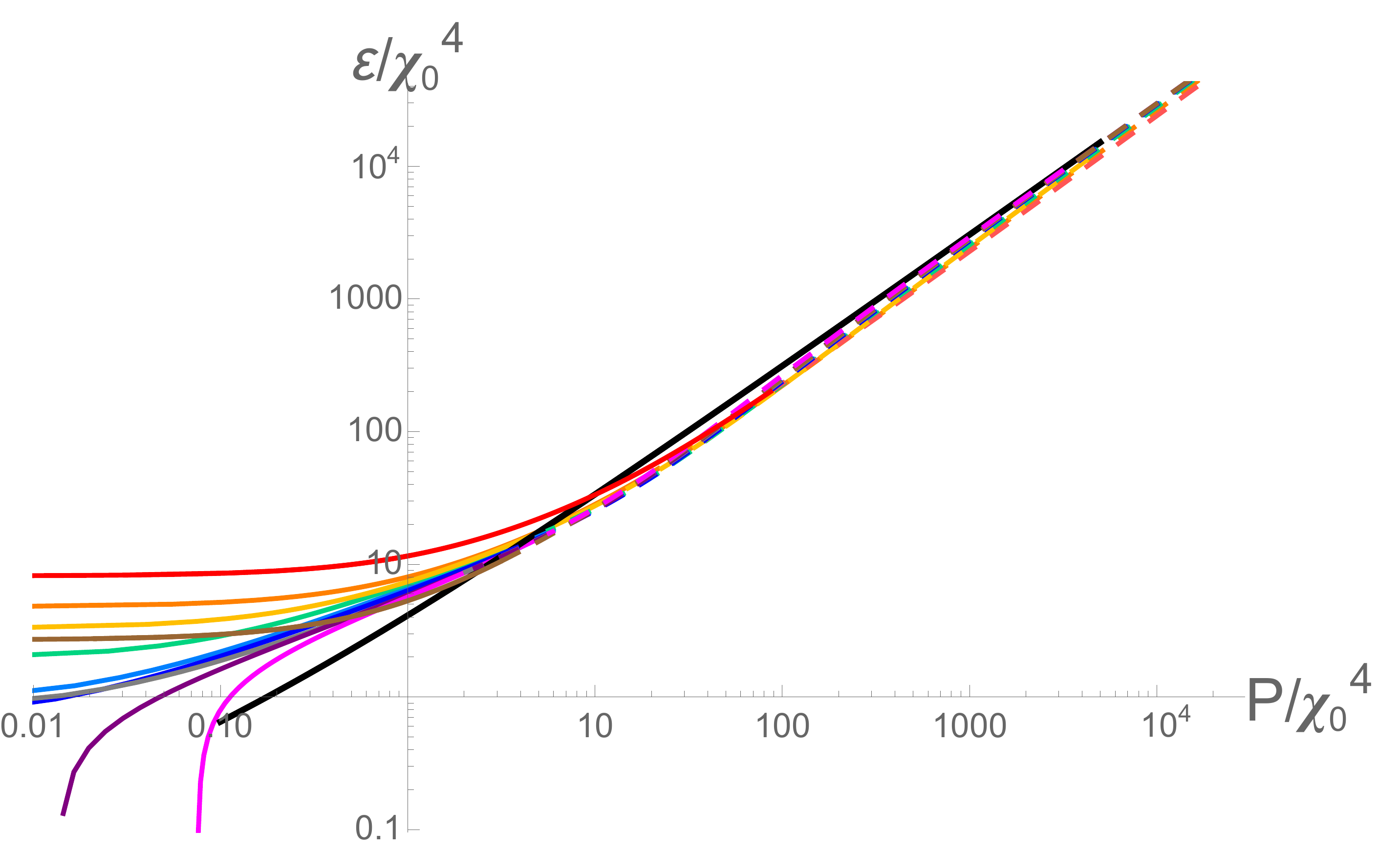}
  \caption{\footnotesize{ \textit{Plots of pressure versus $\mu$ and energy density vs pressure for the holographic model with running anomalous dimension. The coloured lines represent different values of q; (Red) q=1, (orange) q=1.3, (yellow) q=1.45, (green) q=1.6, (light blue) q=1.8, (blue) q=1.99, (gray) q=2.2, (brown) q=2.3, (purple) q=2.4, (magenta) q=2.8 Solid lines are the massive quark phase, dotted lines the chirally symmetric phase. The black lines are the case of a constant dilaton.}}}
            \label{fig:7}
\end{figure}

Next in Fig \ref{fig:7} we plot the pressure (minus the free energy) against $\mu$ for these theories. For each $q$ we mark the lines to show where the novel deconfined yet massive phase and the massless phase are present. Note that in a natural theory, with one scale, in the deep IR we would expect the energy density over $\chi_0$ to lie around one and it does for $q \simeq 2$  - here the theory has the scale where the BF bound is violated very close to one and the derivative of $\gamma$ in the $\rho$ direction at the BF bound violating point is also close to one. Theories with $q$ either much greater than 2, or that are walking, have an extra parameter (the gradient of $\gamma$) that changes the IR behaviour. We include the basic conformal D3/D7 model prediction also (here the phase is massive for all $\mu$). We see that the inclusion of a running anomalous dimension raises the free energy in all cases relative to the basic D3/D7 model - this is to be expected since the dilaton profiles we use increase the action in the IR.    We also show the energy density against pressure to show the theories are all converging in their predictions in the UV whilst distinct in the IR. 

\begin{figure}[]
 \includegraphics[width=7.5cm]{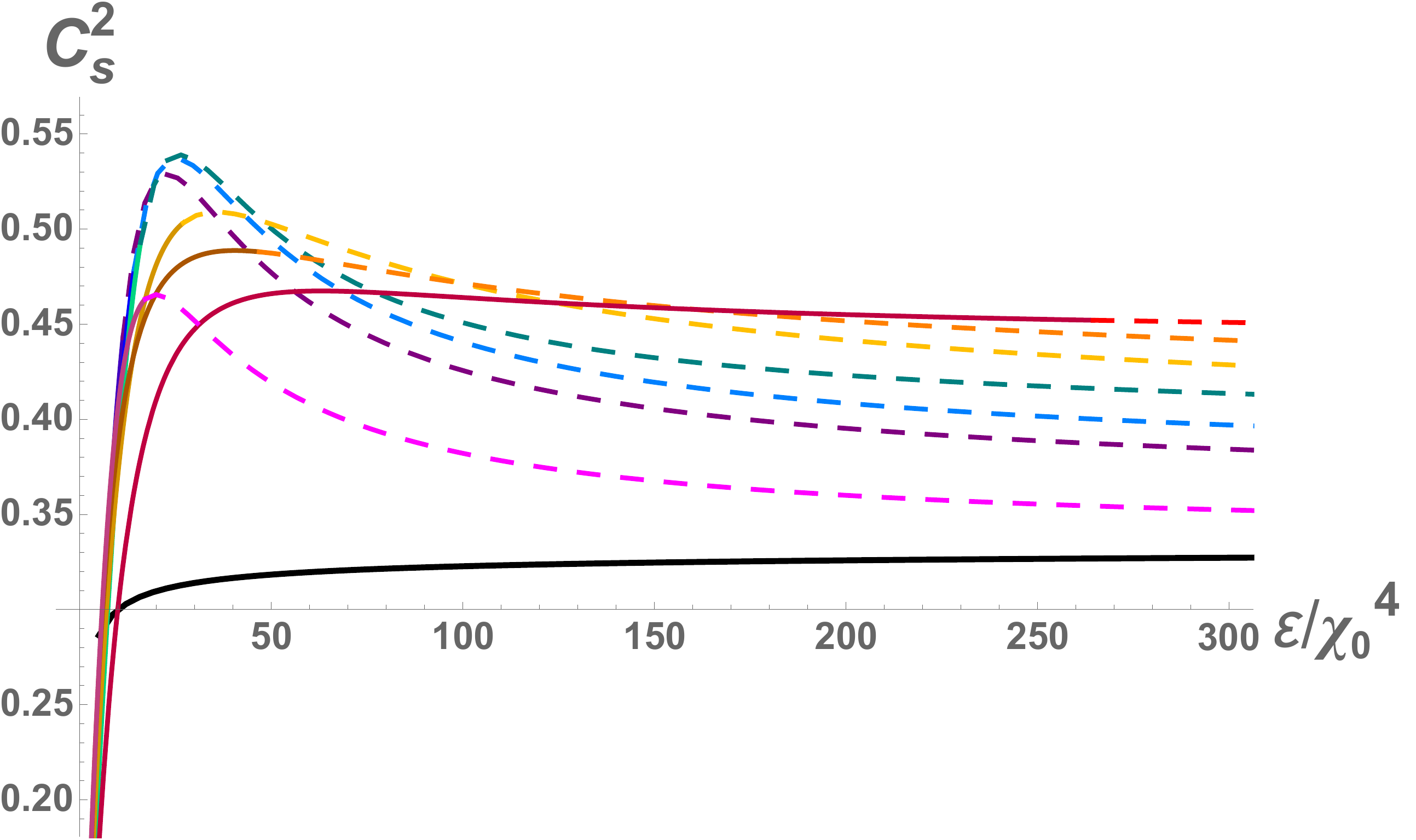} 
  \caption{\footnotesize{ \textit{The speed of sound plotted against energy density in units of $\chi_0$ for theories with different $q$.The coloured lines represent different values of q; (Red) q=1, (orange) q=1.3, (yellow) q=1.45, (green) q=1.6, (blue) q=1.8, (purple) q=1.99, (magenta) q=2.8. Solid lines are the massive quark phase, dotted lines the chirally symmetric phase. The black line is the case of a constant dilaton.}}}
            \label{fig:8}
\end{figure}

The theories with the running anomalous dimension clearly have stiffer equations of state than the basic D3/D7  model and a useful check of how much stiffer is to compute the speed of sound - we show the speed of sound against energy density in Fig \ref{fig:8}. The non-monotonicity of the speed of sound is a notable feature. Here the peak is caused around the scale at which the coupling runs from the UV $\gamma=0$ regime to the IR fixed point regime. This point is also close to the scale where the massive deconfined phase transitions to the chirally symmetric phase. The highest peak seems to occur where in the running of $\gamma$ both the gradient to leave the UV regime and to enter the IR regime are largest.  The higher IR fixed point theories with $q$ just below 2, which naively one would have chosen to represent QCD,  have the highest speed of sound and it rises briefly above 0.5 which is a rough guide to where interesting neutron star physics may occur \cite{Bedaque:2014sqa}- we will investigate this below. Note all the theories asymptote to the speed of sound being a third at high $\mu$. 

\section{Neutron Star Phenomenology}

We have developed a holographic model of the high density regime of QCD with a parameter $q$ that describes a variety of running anomalous dimension profiles. The models include a deconfined yet chirally broken phase and suggest that quite stiff EoS can exist. It's now interesting to see what these models predict for neutron star phenomenology. We first review how to convert our equations of state to a relation between the mass  and radius of a neutron stars.

\subsection{Equations of State and TOV Equations}
The EoS of strongly interacting matter determines the mass-radius relation of neutron stars. This is realized via the Tolman-Oppenheimer-Volkov (TOV) equations 
\begin{eqnarray}
\frac{dP}{dr}&=&-G\left( \mathcal{E}+ P\right)\frac{m+ 4\pi r^3P}{r(r-2Gm)},\\
\frac{dm}{dr}&=&4\pi r^2\mathcal{E} \qquad
\end{eqnarray} 
which are the relativistic equations that model hydrostatic equilibrium inside the stars. $G$ is Newton's constant. Here $m$ and $P$ are the mass and pressure in the star as a function of radius $r$. To integrate the equations we need to input the EoS $\mathcal{E}(P)$, as well as the central pressure $P_c=P(r = 0)$ as initial condition, and the output are the mass $m(r)$ and Pressure $P(r)$ of the corresponding star at a radial distance $r$. The radius $R$ of the star will be the value of $r$ at which the pressure vanishes as we expect outside of the star.  Then varying the initial condition $P_c$ as a parameter we can construct a curve for the mass of the star $M=m(r=R)$ against $R$.

It is useful to place the TOV equations in their dimensionless form:
\begin{eqnarray}
\frac{dp}{d\xi}&=&-B\frac{ye\left(1+\frac{p_0}{\epsilon_0}\frac{p}{e}\right)}{\xi^2(1-2B\frac{p_0}{\epsilon_0}\frac{y}{\xi})}\left(1+A\frac{p_0}{\epsilon_0}\xi^3\frac{p}{y}\right),\\
\frac{dy}{d\xi}&=&A\xi^2e(\xi) \qquad\qquad\qquad
\end{eqnarray} 
Where $r=r_0\xi$, $M=m_0y(\xi)$, $P=p_0p(\xi)$, $\mathcal{E}=\epsilon_0e(\xi)$, $A=\frac{4\pi r^3_0 \epsilon_0}{m_0}$ and  $B=\frac{Gm_0\epsilon_0}{p_0r_0}$.

We will fix the scale with the value of $p_0=\epsilon_0=\frac{(308.55 MeV)^4}{\pi^2}$ as is sensible in the context of the nuclear equation of state discussed above; this choice then fixes the rest of our scale parameters.

One can make a radial perturbation of a solution. In terms of  the mass vs radius curve one increases the value of the central density $\mathcal{E}_c$ whilst keeping the same mass. If $\frac{\partial M(\mathcal{E}_c)}{\partial\mathcal{E}_c} >0$ then the corresponding equilibrium solution for this new configuration has a higher mass and  therefore there is a deficit of mass. The gravitational force thus needs to be balanced by increasing the central pressure. The forces acting on the matter in the star will therefore act to return the new configuration toward its original unperturbed place. However for the case in which $\frac{\partial M(\mathcal{E}_c)}{\partial\mathcal{E}_c)} \leq 0$ we arrive at the conclusion that, if the star is perturbed, the forces acting on the perturbed star will act to drive it further from its original point in the mass vs radius curve. Therefore the  condition for stability is given by  (\ref{stability}). As mentioned in \cite{Alford:2017vca} we can also determine the stability of a star from the mass vs radius curve using the Bardeen, Thorne and Meltzer (BTM) criteria \cite{BardeenThorneMeltzer} which established a simple formulation to know if all its radial modes are stable:

\begin{enumerate}
\item At each extremum where the $M(R)$ curve  rotates counter-clockwise with increasing central pressure, one radial stable mode becomes unstable.
\item  At each extremum where the $M(R)$ curve  rotates clockwise with increasing central pressure, one unstable radial stable mode becomes stable.
\end{enumerate}

\subsection{Mass Radius Relations}

\subsubsection{Nuclear phase}

In section IIA we included three equations of state from \cite{Hebeler:2013nza} for the nuclear phase above 308.55 MeV.

\begin{figure}[h]
 \includegraphics[width=8cm]{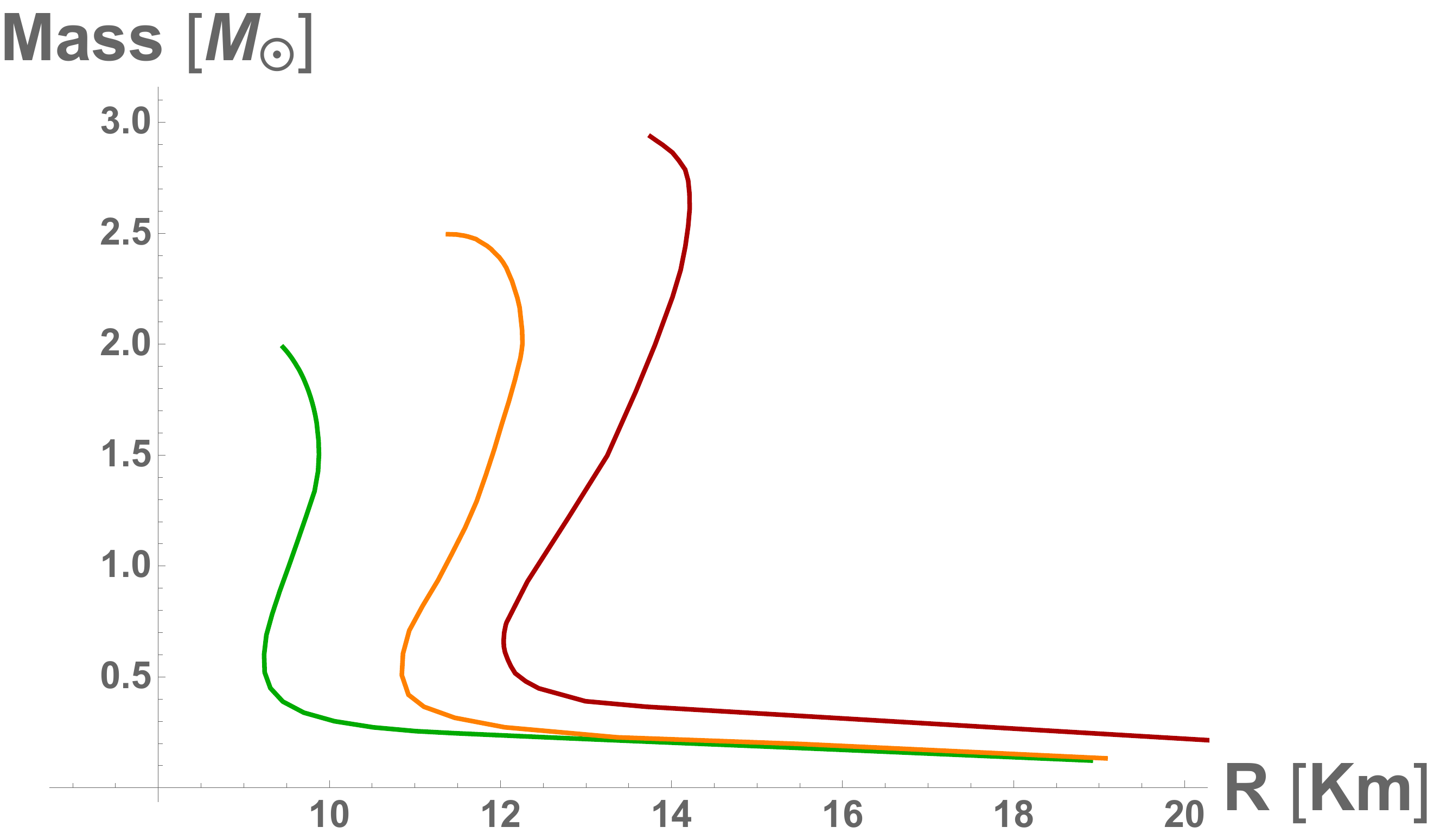}  
  \caption{\footnotesize{ \textit{Mass of the Neutron Star (in units of solar mass $M_{\odot}$) as a function of its radius (in kilometers) for nuclear matter from EFT EoS. The Green line represents a soft EoS, the orange a medium EoS and the red line a stiff EoS.}}}
  \label{fig:9}
\end{figure}

To obtain the mass vs radius curve  we solve the TOV equations starting from the highest density region (center of the star), using the numerical equation of state.  The maximum density the equations of state are consistent for (see section IIA) set a maximum neutron star mass in each case. The result of the computations, confirming previous analysis is shown in Figure \ref{fig:9}. The observation of neutron stars in the 2-2.5 solar mass range suggest that the stiffer EoS are more physical. 

\begin{figure}[h]

 \includegraphics[width=8cm]{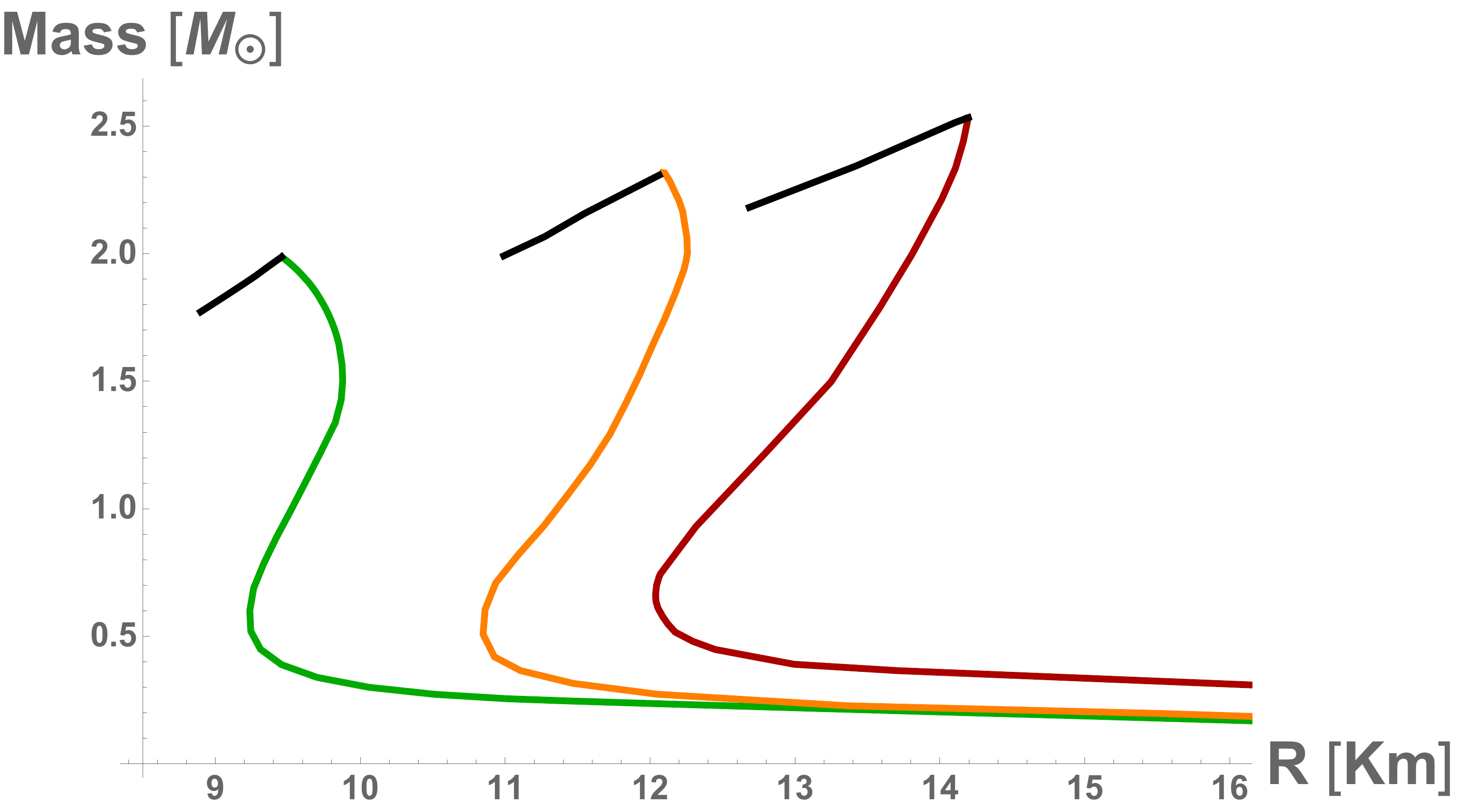}  
  \caption{\footnotesize{ \textit{Mass of the Neutron Star  (in units of solar mass $M_{\odot}$) as a function of its radius (in kilometers). Colour lines represent Nuclear matter star from EFT EoS, the black lines represent the change of phase towards a hybrid star with a quark core using the constant dilaton D3/D7 model.}}}
  \label{fig:10}
\end{figure}

\subsubsection{Basic D3/D7}

As a further cross check of our methods we reproduce the mass radius plot for neutron stars with the equation of state from section IIB1. That is the basic, constant dilaton D3/D7 model of \cite{Karch:2007br} with the mass scale set so that the transition for the on-set of density occurs at $\mu=$ 308.55 MeV. The transitions to the high density phase are those shown in Figure \ref{fig:10}. As in \cite{Hoyos:2016zke}  we find only unstable stars with a core of this material.

\begin{figure}[h]
\includegraphics[width=8cm]{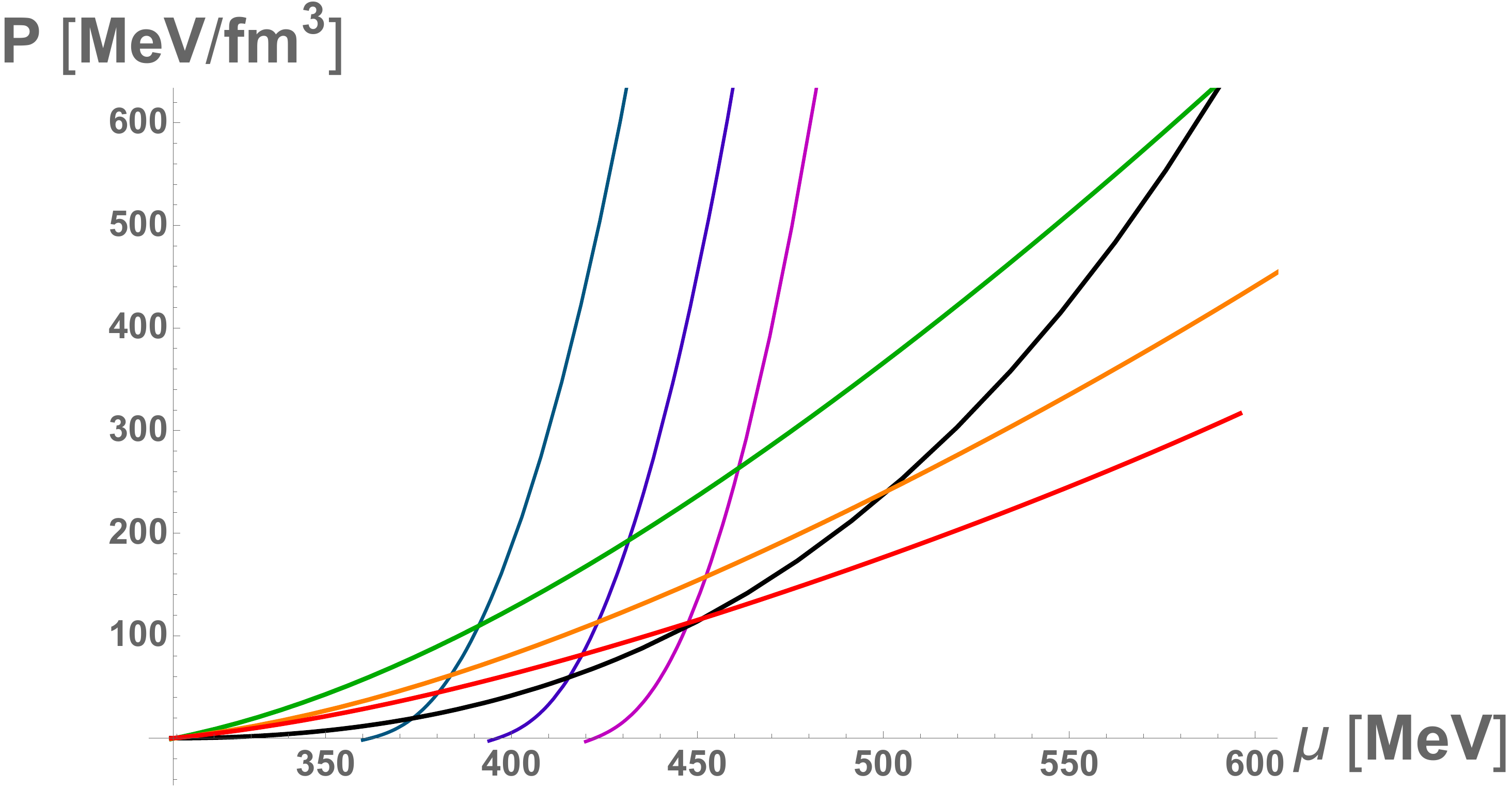}    
  \caption{\footnotesize{ \textit{ Transition from nuclear to quark matter for the case of q=1.8. The Black line correspond to the case of a constant dilaton and the green, orange and red curves represent nuclear matter as in Fig \ref{fig:4}. The dark teal curve corresponds to $\chi_0=360$ MeV, the purple curve corresponds to $\chi_0=395$ and the magenta curve corresponds to $\chi_0=420$. }}}
  \label{fig:11}
\end{figure}

\subsubsection{Bottom-Up D3/D7 with Running $\gamma$}

We have seen that our bottom up models have a stiffer equation of state when the running anomalous dimension of the quarks is included. In fact, as we will see, only the stiffest models with $c_s^2 > 0.5$ are of any interest phenomenologically for neutron stars. Let us therefore begin by studying the case $q=1.8$ which has the stiffest equation of state.

For $q=1.8$ we must also pick the scale $\chi_0$. Naively this is roughly 330 MeV (a third the proton mass) but if we make such a low choice the nuclear phase barely exists before the quark phase takes over. The naive relation to the proton mass though is only an estimate so we will allow ourselves to consider a range of test cases: $\chi_0=360, 395$ and $420$ MeV. In Fig \ref{fig:11} we show the pressure against chemical potential plots for these cases - the nuclear curves are also displayed so the position of the phase transitions can be read off. Note the transition to the quark phase are typically at lower scales than in the basic D3/D7 model since the pressure is larger. 

\begin{figure}[h]
\includegraphics[width=8cm]{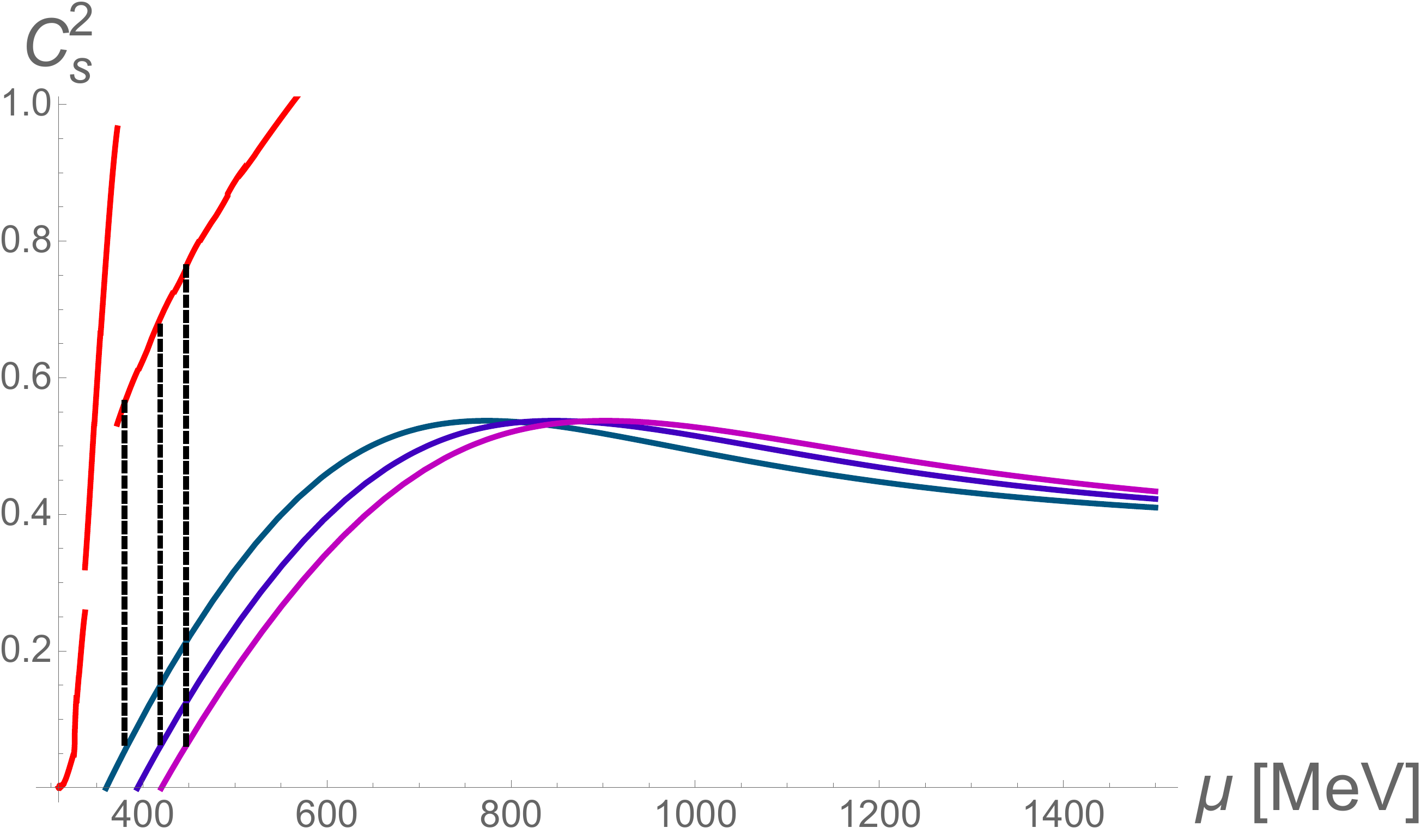}  
 \includegraphics[width=8cm]{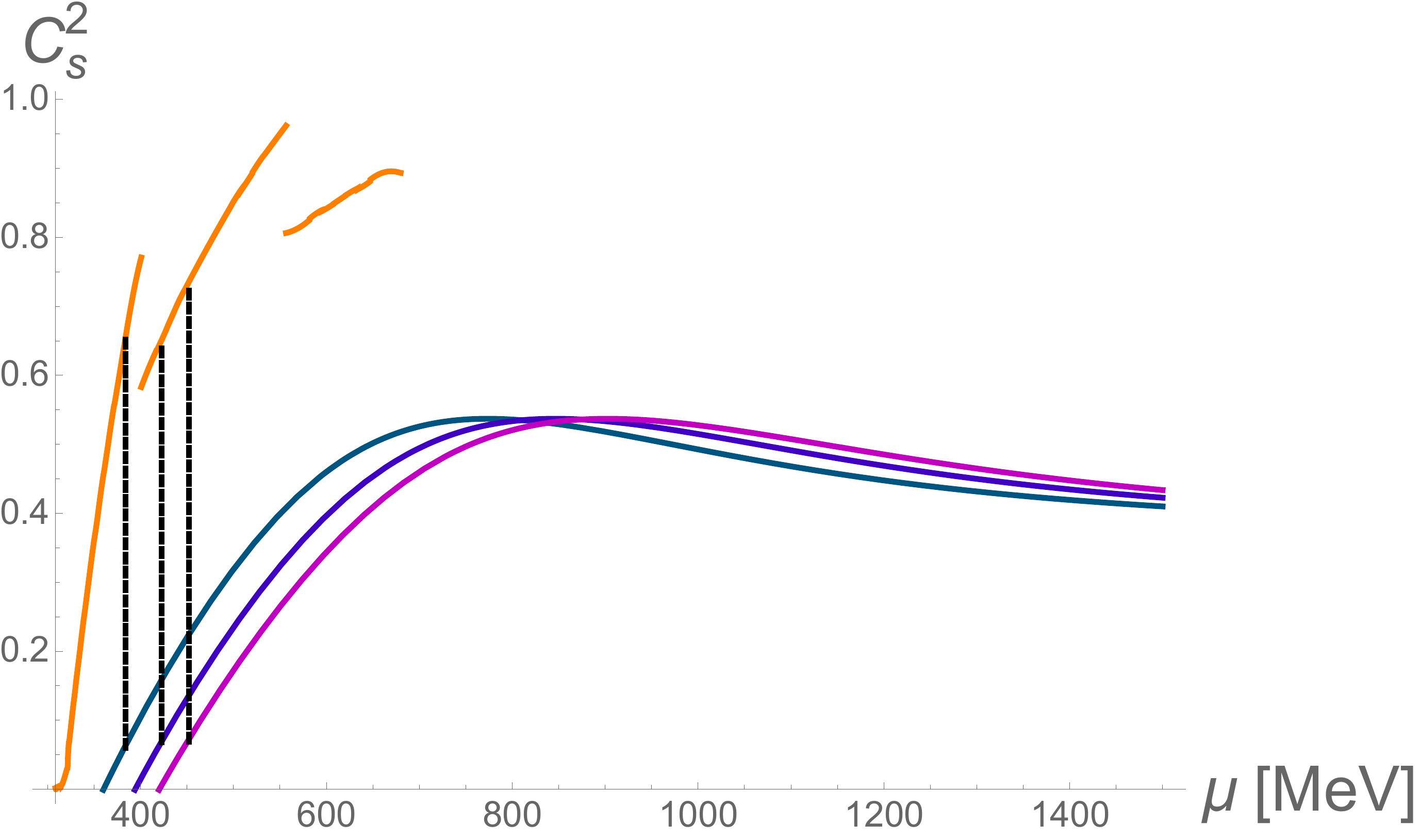} 
  \includegraphics[width=8cm]{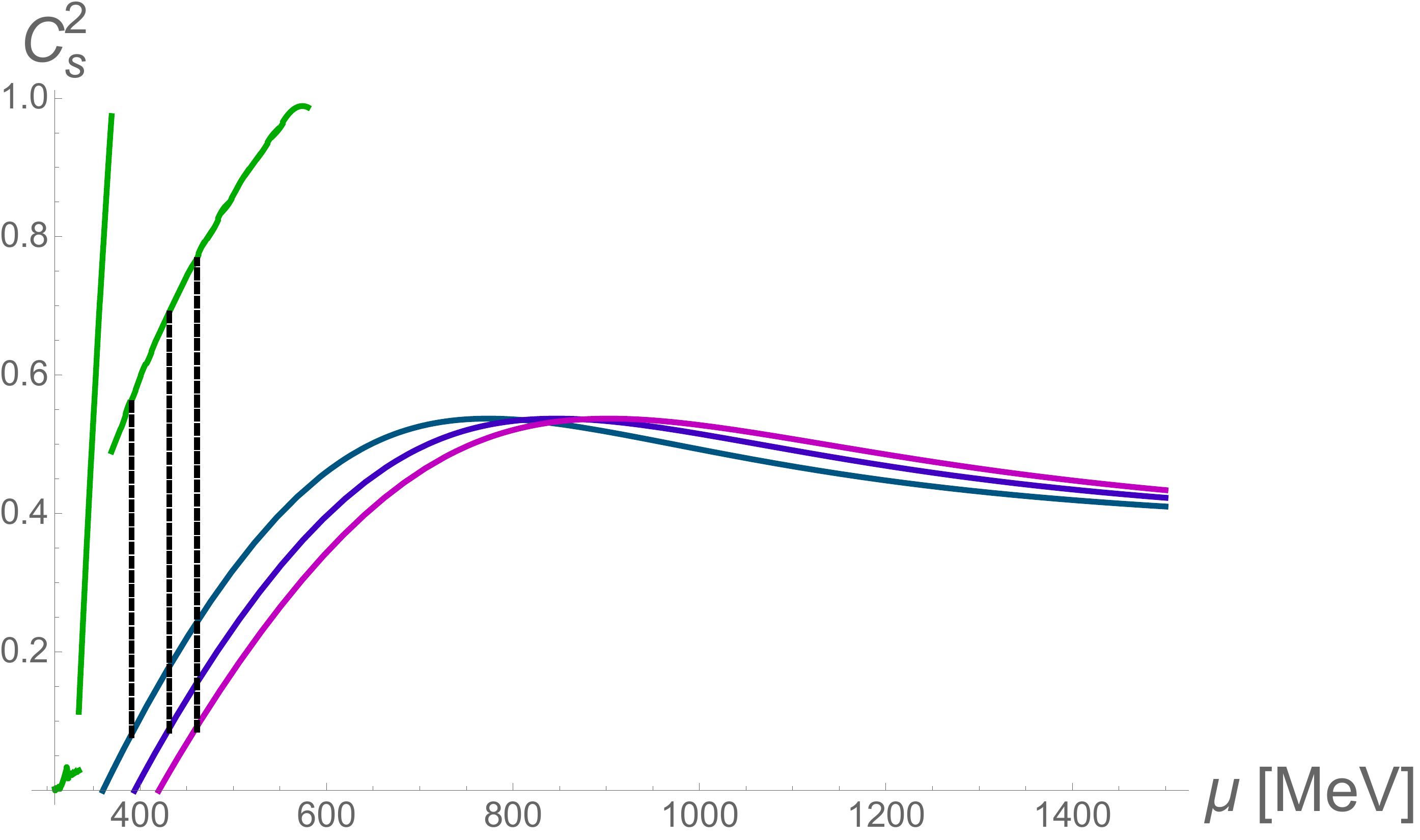}   
  \caption{\footnotesize{ \textit{ Speed of sound squared as a function of the chemical potential for the case of q=1.8. Green, orange and red curves are those for the three nuclear EoS and the three quark matter curves are in dark teal for $\chi_0= 360$ MeV, in purple for $\chi_0= 395$ MeV, and in magenta for $\chi_0=420$ MeV. The transition from nuclear to quark matter is indicated with a black dashed line.}}}
  \label{fig:12}
\end{figure}

It is instructive to see how stiff the quark matter is at the transition. In Fig \ref{fig:12} we plot $c_s^2$ against $\mu$ separately for each of the nuclear equations of states. The black dotted lines show  where the phase transitions occur. Clearly there is a distinct drop in $c_s^2$ as one moves to the quark phase in all these cases but the stiffness does then grow at higher $\mu$. One might expect that the neutron star stability will decay when the core moves above the transition but that there might be a new class of stars with the denser cores reflecting the stiffness at higher $\mu$.

\begin{figure}[h]
\includegraphics[width=8cm]{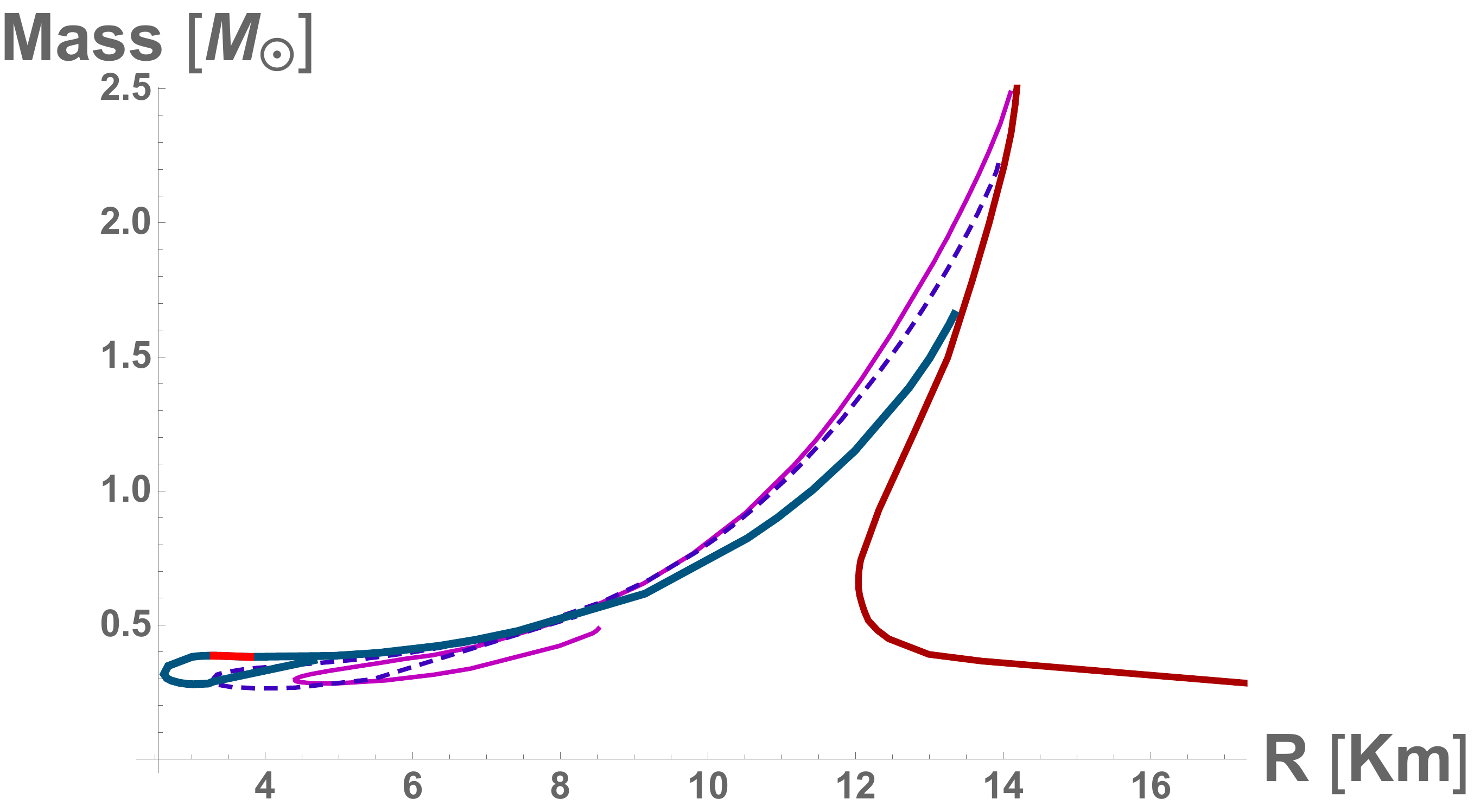}  
 \includegraphics[width=8cm]{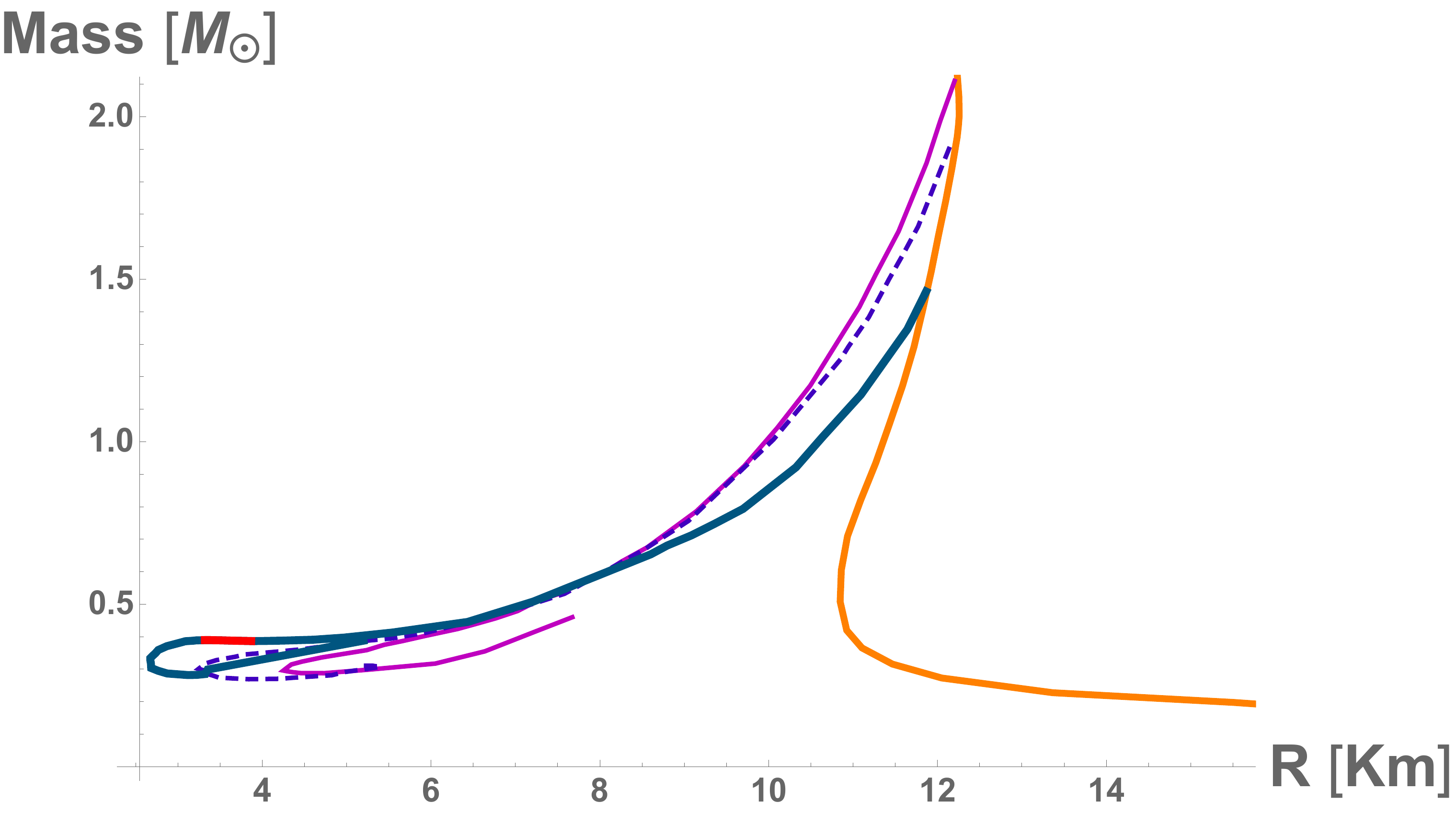} 
  \includegraphics[width=8cm]{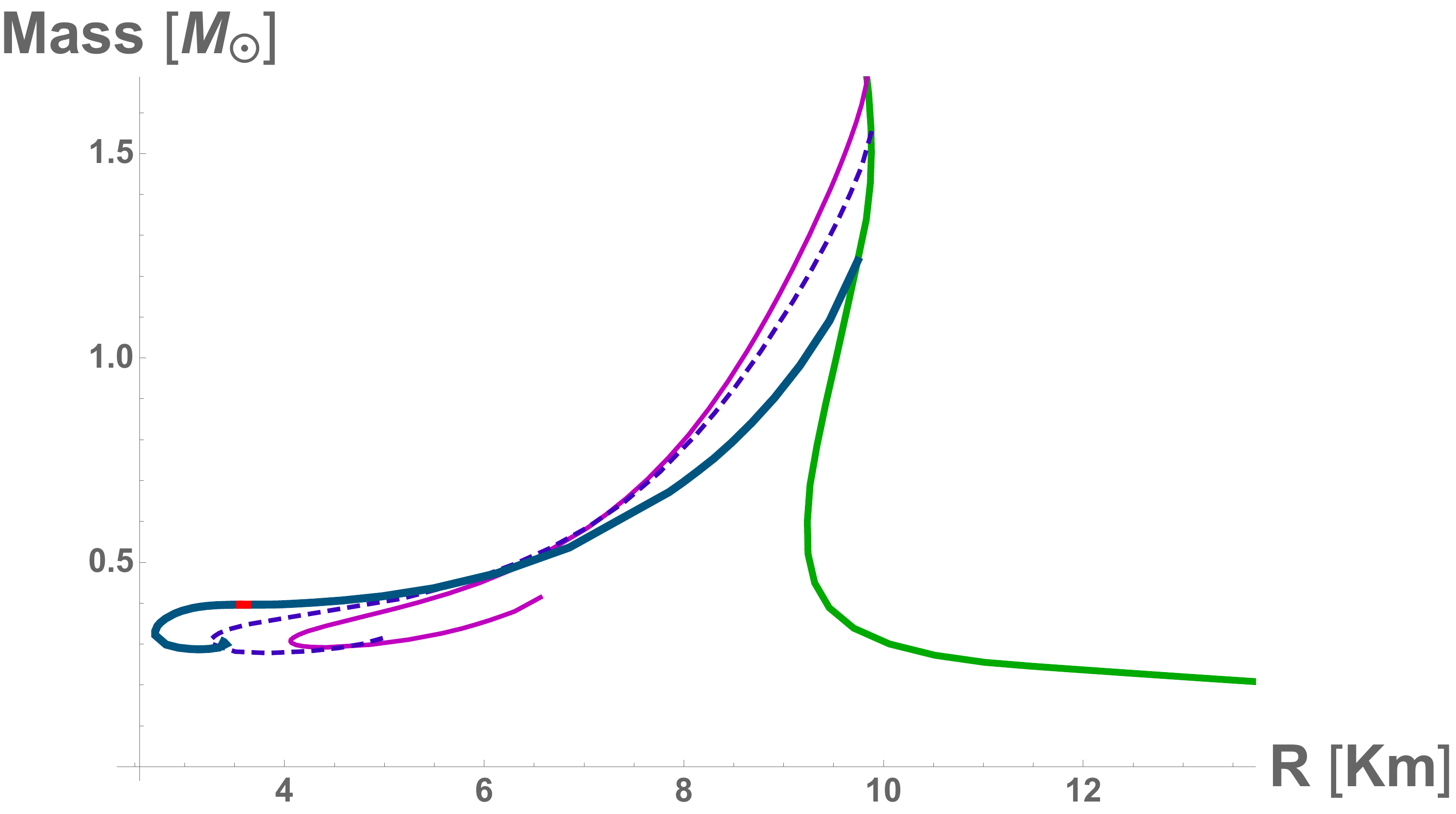}   
  \caption{\footnotesize{ \textit{ Mass vs radius curves for the case of q=1.8. The three curves leaving the green/red/orange nuclear EoS prediction are the three transitions to a quark phase from Figure \ref{fig:12}. The small stable branch is indicated in red.}}}
  \label{fig:13}
\end{figure}

We solve the TOV equations for these cases and display the mass vs radius curves in Fig \ref{fig:13}. The results indeed fit our intuition. The stable neutron star branch ends in all cases when the transition to the quark matter occurs. The stiff area of the equation of state does kick in again though hinting at a new branch of smaller, lighter, hybrid stars with quark matter cores - the stable solutions are marked in red. Only for the softest nuclear equation of state are there, briefly, truly  stable hybrid stars with quark matter cores  but clearly in all cases the EoS is close to stiff enough to make such solutions. We should also immediately caveat that such a disconnected branch of smaller stable stars might not be possible to produce astrophysically. On the other hand in the light of the expected new gravitational wave data it is interesting to find exotics in case there are surprising signals. Note in our model in no case are there both quark core hybrid stars and neutron stars as massive as 2 solar masses.  Nevertheless the solutions suggest that our EoS are closer than those in \cite{Hoyos:2016zke}  to generating interesting phenomenology and perhaps with a further increase in stiffness of the EoS both could be realized.

\begin{figure}[h]

\includegraphics[width=8cm]{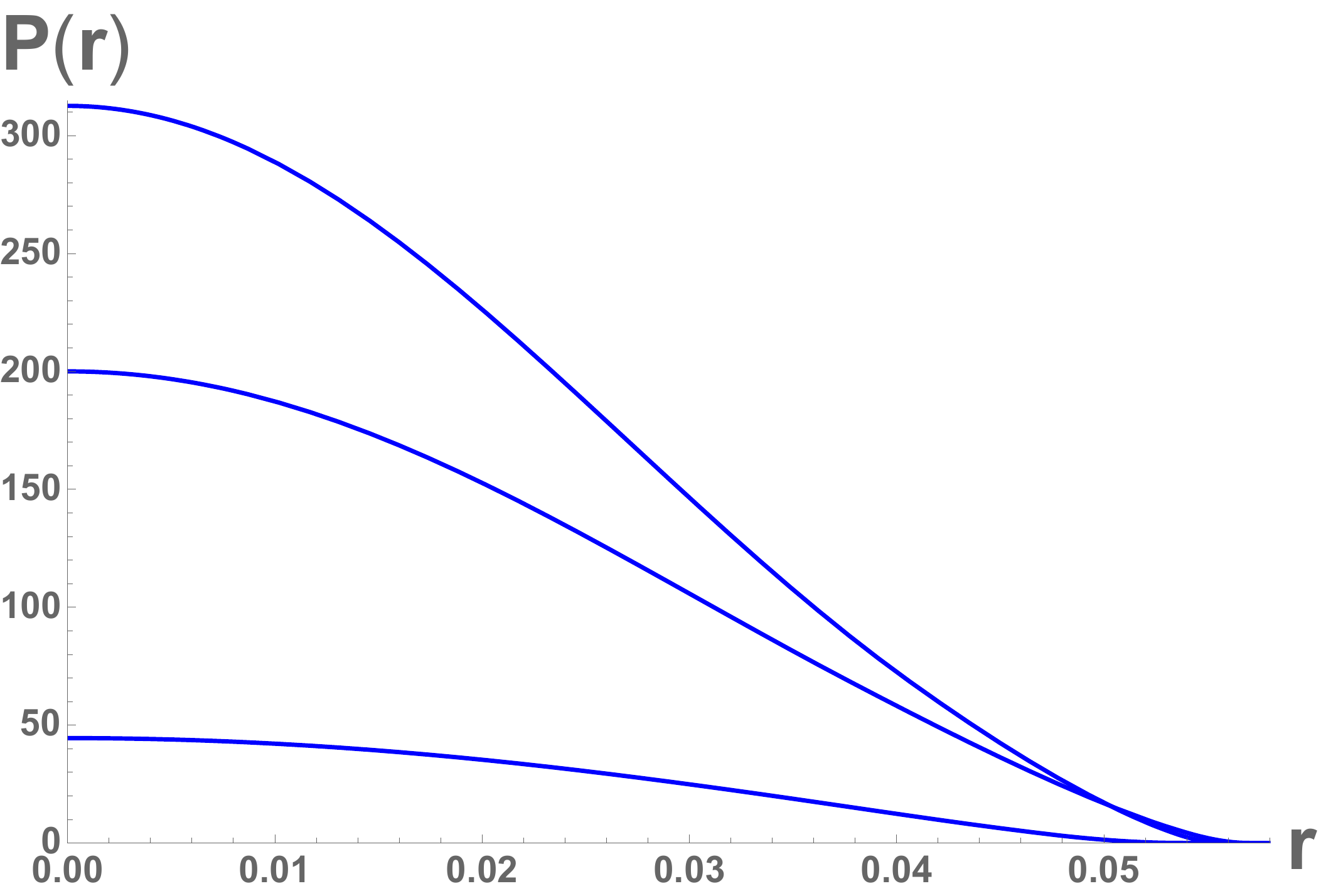}
  
(a)

 \includegraphics[width=8cm]{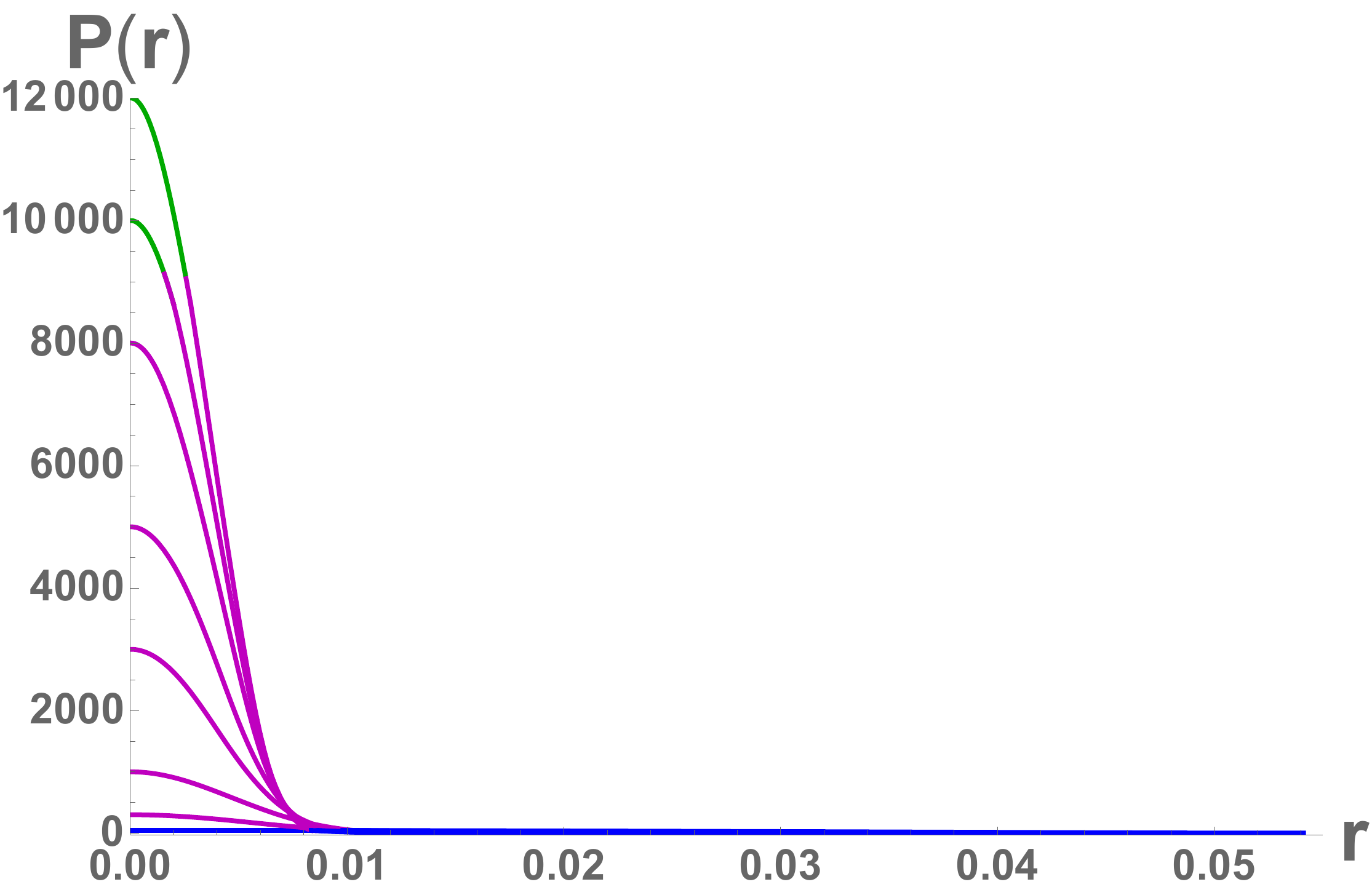}  
 
 (b)
  \caption{\footnotesize{ \textit{ Pressure as a function of the radial variable r. The radius of the Neutron star is the value of r at which P(r) vanishes. (a) Pressure for the case of stiff nuclear matter taken from reference \cite{Hebeler:2013nza} (b) Pressure for a hybrid star where the quark phase (the pink line corresponds to the massive chirally broken phase and the green line corresponds to the massless chirally symmetric phase) correspond to a value of q=1.8 and $\chi_0=360 MeV$. Note the stable cases from Fig \ref{fig:13} lie where the chirally symmetric phase just enters at the center and the speed of sound is highest (see Fig \ref{fig:8}).  }}}
  \label{fig:14}
\end{figure}

It is interesting to understand the difference in composition of the traditional neutron stars and the new class of stable stars we are predicting here. In Figure \ref{fig:14} we plot the pressure against radius in representative stars with the different phases distinguished. Note the neutron stars have very different central pressures for very similar radii reflecting the sharp rise in speed of sound/stiffness of the neutron equations of state needed to support 2 solar mass neutron stars.  The novel hybrid stars are very much quark matter dominated and rely on a broader softer core for stability.

These results have been for the case $q=1.8$ which has the stiffest EoS and highest peak speed of sound. Lower or higher $q$ values have softer EoS and produce no new conclusions beyond the instability of the hybrid stars. We do not therefore present any analysis of those cases. 

The EoS in the improved holographic models are still not stiff enough to play a role in compact object phenomenology although the equations hint that they may be closer to a role than those proposed in \cite{Hoyos:2016zke} . This suggests further refinements may lead to interesting predictions.

\subsection{Restoring Confinement}

Our equations of state so far either don't support hybrid stars or are at odds with the 2 solar mass neutron star observations. This need not be the final conclusion though. We have modified the D3/D7 model (which in base form has neither confinement nor chiral symmetry breaking) to include chiral symmetry breaking. We have not though included confinement. 

A justification for this is that chiral symmetry breaking may well set in before confinement. The QCD coupling might run to a critical value for chiral symmetry breaking at which scale the quarks will become massive and decouple from the pure Yang Mills theory running. That running is very fast and starting at rather strong coupling and will very quickly reach any critical value for confinement in the pure glue theory so that confinement and chiral symmetry breaking are intimately linked and lie very close in scale. The D3/D7 system we have does not include this change in phase to confined though and so only describes the phases above the deconfinement transition fully. 

The main impact of this omission is that we may be wrongly computing the vacuum energy of the $\mu=0$ phase of QCD by a constant factor. Then we are placing the phase transitions in the wrong place. We have explored adding such a ``bag constant" factor. 

\begin{figure}[h]
\includegraphics[width=8cm]{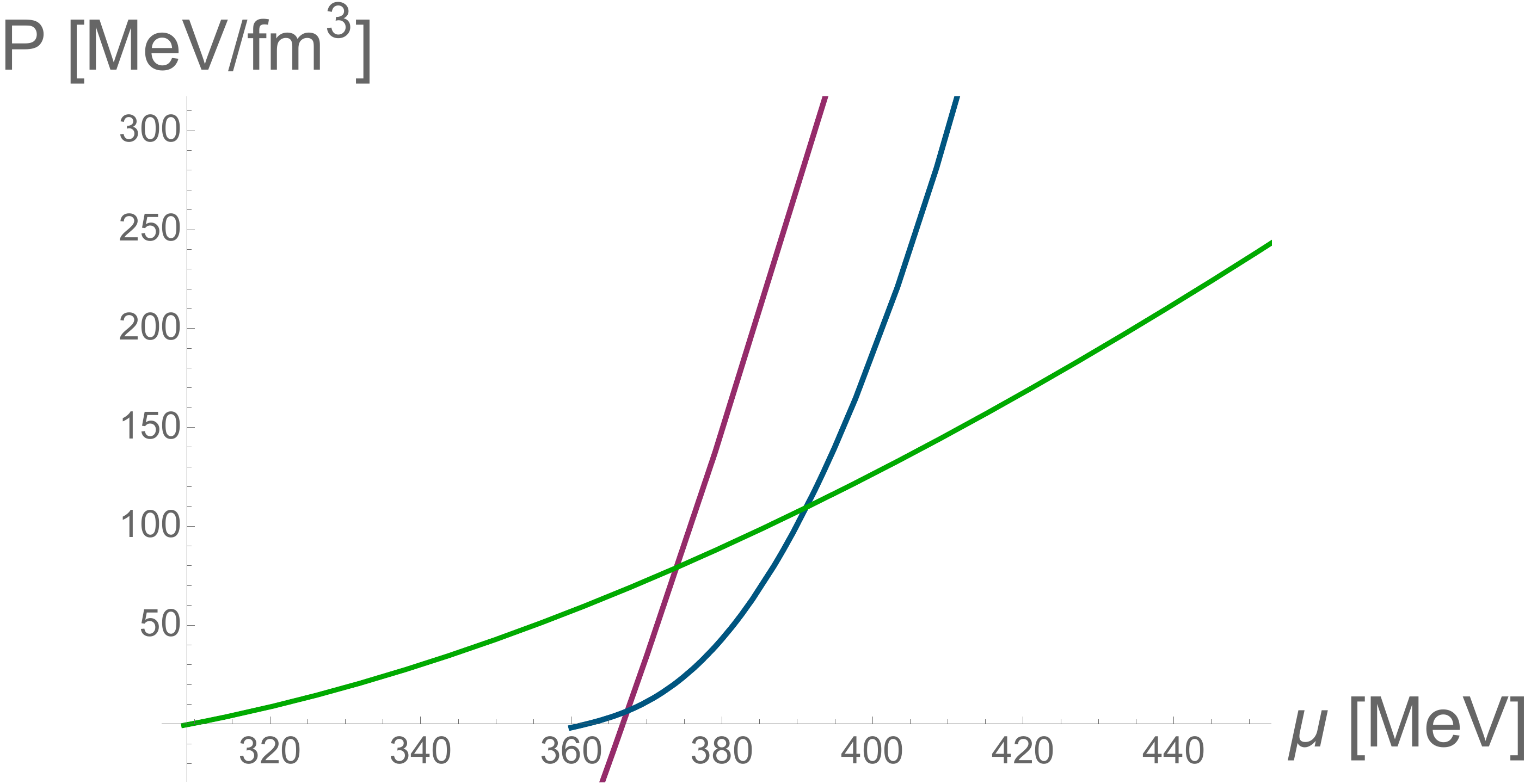}  

(a)

 \includegraphics[width=8cm]{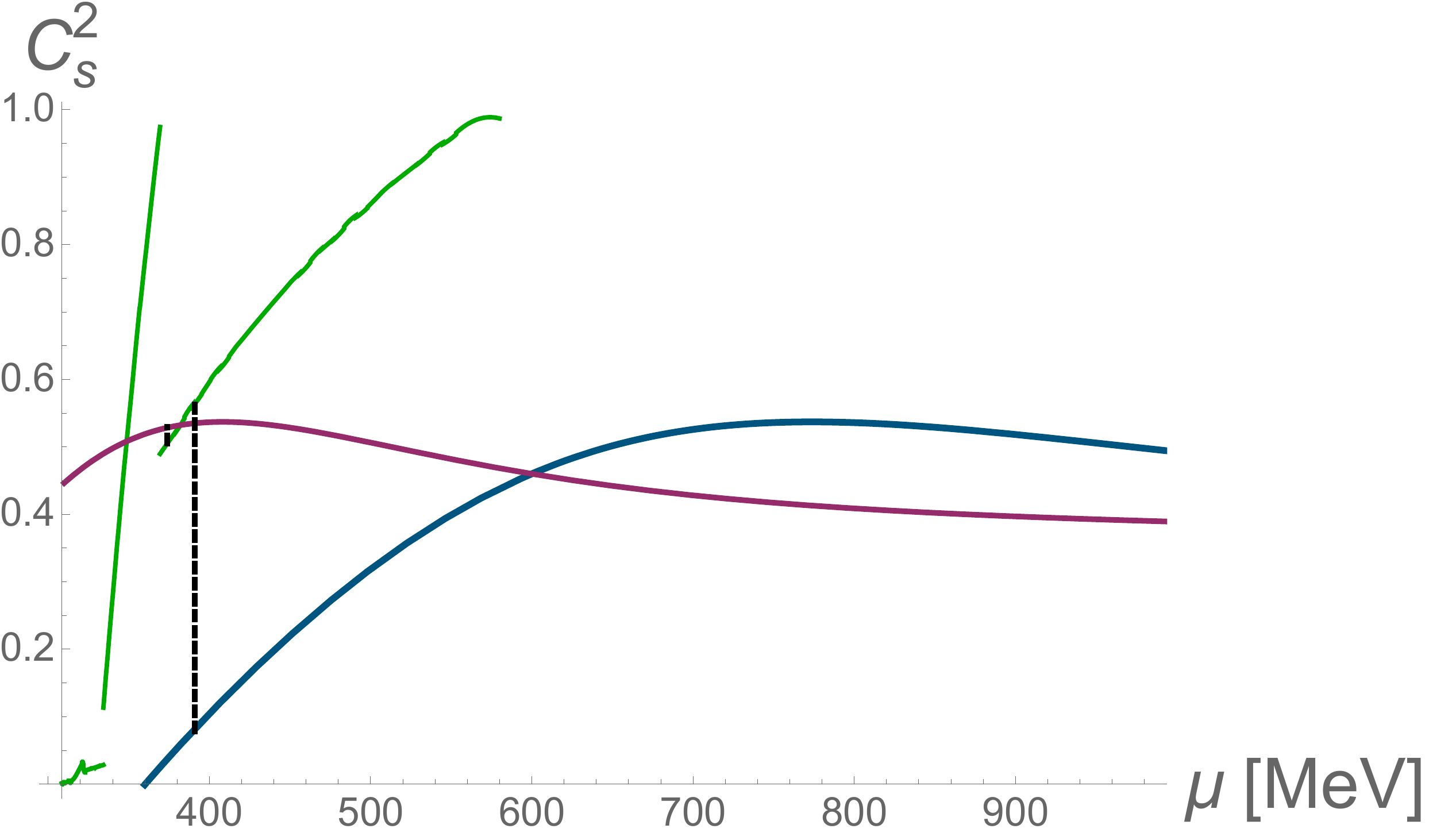}  
 
 (b)
 
  \includegraphics[width=8cm]{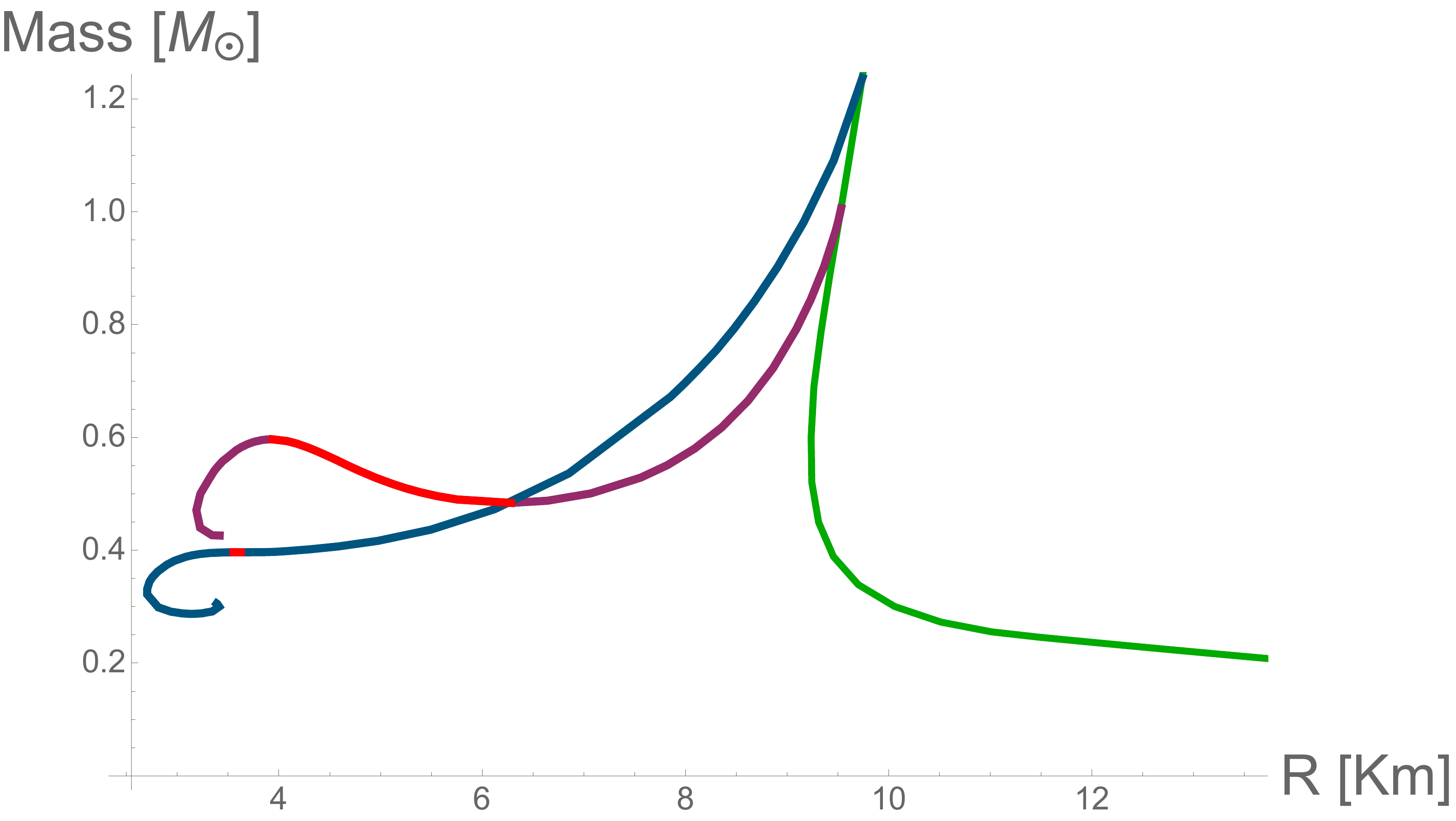}  
 
 (c)
  \caption{\footnotesize{ \textit{(a) Pressure vs chemical potential for different phases. The nuclear phase (green) correspond to soft nuclear matter; the quark phases: (purple) correspond to a value of $q=1.8$ and $\chi_0=190 MeV$ and (dark teal) correspond to a value of q=1.8 and $\chi_0=360 MeV$. (b) Comparison of the speed of sound squared in units of c as a function of the chemical potential for the same phases. We show with a black dashed line the point of transition between the nuclear phase and the quark phase. (c) Mass vs radius curve showing the same phases as above. The stable branches are indicated in red. }}}
  \label{fig:15}
\end{figure}

The subtraction of such a constant from the high energy phase  free energy allows us to set $\chi_0$ smaller than previously whilst maintaining a low density nuclear phase. We can then move the region of $\mu$ where the high density phase has a large speed of sound  closer to the transition point. Generically though we have not been able to maintain the neutron star branch of stable stars with ones with quark cores - the quark matter transition always leads to the neutron star branch being unstable (before a 2 solar mass neutron star is achieved). We can though make the novel hybrid stars we have seen  more stable in this way. In Figure \ref{fig:15} we show an example of the most sympathetic case with a substantial hybrid  star region.

\section{Conclusions}

The existence of neutron stars up to and over 2 solar masses provides a challenge in our understanding of the QCD equation of state (EoS) even within nuclear matter models. At the cores of these stars it seems the matter must be very stiff with speeds of sound close to the speed of light. Gravitational wave signals from colliding neutron star pairs are also beginning to constrain the EoS through measurements of the tidal deformability. It is therefore interesting to study the deconfined quark matter equations of state to see if they might play a role in the cores of neutron stars or generate other hybrid stars. This requires knowledge of and the ability to calculate in the strongly coupled yet deconfined section of the QCD phase diagram. There are no first principles tools that can be brought to bare since the lattice can not compute at sizeable chemical potential. This motivates trying  to use holography to explore possible descriptions of this regime in QCD.

The first holography paper addressing neutron star structure \cite{Hoyos:2016zke}  used the exact results at finite $\mu$ for the D3/D7 dual system. That system though has conformal gauge dynamics and no chiral symmetry breaking unless introduced by a hard mass. It predicted a very soft equation of state that could not play a role in neutron star phenomenology. Our goal here has been to adjust that model to include a running anomalous dimension for the quark condensate which introduces a dynamical chiral symmetry breaking mechanism. Such theories suggest a massive deconfined phase with deconfined quarks yet chiral symmetry breaking before moving to the chirally restored high density phase. We have shown that this leads to a stiffer equation of state in the relevant intermediate $\mu$ phase and that the speed of sound has the required rise and fall (see the non-monotonicity in Figure \ref{fig:8}) in this regime. 

We have used the TOV equations to model compact stars using our EoS varying the IR quark mass. The instability of the neutron star branch remains but in some case we do see novel hybrid stars with quark matter cores form. The models hint therefore at twin stars -  two classes of 0.5 solar mass object with very different radii. Naively one supposes that these stars will not be produced in astrophyiscal processes but they are a possible exotic signature in gravitational wave data. Our model does not produce a sufficiently high speed of sound in the material to allow both 2 solar mass neutron stars and hybrids to exist together although the EoS are clearly close to realizing this.  They also do not support a branch of solutions beyond the standard neutron star branch that link continuously to neutron stars. Nevertheless, we view this work as the next step beyond \cite{Hoyos:2016zke} towards a full model. In the future models that do a better job of including confinement and colour superconducting phases may be possible and yet stiffer EoS may emerge.

\bigskip \noindent {\bf Acknowledgements:}  We thank A. Schmitt for useful discussions. 
NE's work was supported by the
STFC consolidated grant ST/P000711/1 and JCR's by Mexico's National Council of Science and Technology (CONACyT) grant 439332. KBF would like to acknowledge the financial support of Shahrood University of Technology for this research under project No 24054. 


\end{document}